\documentclass[twocolumn]{aastex61}

\received{March 11, 2019}
\revised{May 16, 2019}
\submitjournal{AJ}
\shorttitle{Oort Planets}
\shortauthors{Bailey \& Fabrycky}

\usepackage{graphicx}
\begin{document}

\title{Stellar Flybys Interrupting Planet-Planet Scattering Generates Oort Planets }

\author[0000-0001-7509-0563]{Nora Bailey}
\correspondingauthor{Nora Bailey}
\email{norabailey@uchicago.edu}

\author[0000-0003-3750-0183]{Daniel Fabrycky}
\affil{Department of Astronomy \& Astrophysics, University of Chicago, Chicago, IL 60637}

\begin{abstract}
Wide-orbit exoplanets are starting to be detected, and planetary formation models are under development to understand their properties. We propose a population of ``Oort'' planets around other stars, forming by a mechanism analogous to how the Solar System's Oort cloud of comets was populated. Gravitational scattering among planets is inferred from the eccentricity distribution of gas-giant exoplanets measured by the Doppler technique. This scattering is thought to commence while the protoplanetary disk is dissipating, $10^6-10^7$~yr after formation of the star, or perhaps soon thereafter, when the majority of stars are expected to be part of a natal cluster. Previous calculations of planet-planet scattering around isolated stars have one or more planets spending $10^4-10^7$~yr at distances $>$100~AU before ultimately being ejected. During that time, a close flyby of another star in the cluster may dynamically lift the periastron of the planet, ending further scattering with the inner planets. We present numerical simulations demonstrating this mechanism as well as an analysis of the efficiency. We estimate an occurrence of planets between 100 and 5000 AU by this mechanism to be $<$1\% for gas giants and up to a few percent for Neptunes and super-Earths. 
\end{abstract}


\section{Introduction} \label{sec:intro}

The outer parts of planetary systems are still mysterious. Myth and speculation shroud the outer parts of our own system, both in the popular imagination and in scientific circles. From Nirabu, to a companion triggering comet showers \citep{1984Whitmire}, to Planet X perturbing the outer planets (\citealt{1988Seidelmann}, but see \citealt{1993Standish}), the unknown is intriguing.

Around the Sun, searches by optical light have come up empty \citep{2010Schwamb,2015Brown}. Gravitational probes have been developed \citep{2005Gaudi,2005Zakamska,2016Fienga}, but have found no convincing evidence. The latest motivation for such searches, the clustering properties of high-perihelion  objects \citep{2014Trujillo,2016Batygin} or collective inclination of the Kuiper Belt \citep{2017Volk}, have touched off renewed efforts to find such planets. Theories of migration of the gas giants suggest one or more ice giants could have been ejected \citep{2012Batygin,2018Nesvorny}, although maybe a planet was instead placed in the Oort cloud. 

In other planetary systems, we have good limits on the frequency of gas giant planets from direct imaging from 30 AU to about 300 AU, with an overall occurrence rate of 5-13 $M_J$ companions of $0.6^{+0.7}_{-0.5} \%$. For FGK stars, there is an upper limit occurrence rate of $5.1\%$ for companions between 100 AU and 1000 AU \citep{2016Bowler}. This range to 1000 AU was also studied via a survey with Spitzer Space Telescope \citep{2016Durkan} that imaged the space around white dwarfs and found, ``assuming a mass distribution of ${dn}/{dm}\propto {m}^{-1.31}$, we constrain (at 95\% confidence) a population of 0.5-13 $M_J$ planets at separations of 100-1000 AU with an upper frequency limit of 9\%.'' The Wide-orbit Exoplanet search with InfraRed Direct imaging (WEIRD) survey \citep{2018Baron} extended this range out to 5000 AU, finding a constraint on ``the occurrence of $1-13 M_J$ planetary-mass companions on orbits with a semi-major axis between 1000 and 5000 AU at less than 0.03, with a 95\% confidence level.'' While direct imaging surveys provide mostly upper limit constraints on wide-orbit companions, the NASA Exoplanet Archive lists 10 confirmed planets beyond 100 AU with mass $<$13 $M_J$ (as of April 29, 2019), implying a non-zero lower limit to wide-orbit planets.

Theoretically, we expect planets to be able to form \emph{in situ} only out to $\sim$30 AU by core accretion \citep{2010Murray-Clay}. Formation could perhaps occur beyond 100 AU out to several hundred AU by disk instability (limited mainly by the observed sizes of protoplanetary disks; \citealt{2009Andrews}). However, such planets would form in high-mass disks and would be quite massive ($\sim$10 M$_J$; \citealt{2005Rafikov}) and are likely to migrate rapidly inward \citep{2011Baruteau}. An analysis involving photoevaporation of the disk showed that migration may be directed outward \citep{2004Veras}, but even this mechanism would not populate the region beyond several hundred AU.

To our knowledge, only one mechanism has been proposed that can place planets on stable orbits of $\sim$1000 AU: a free-floating planet may bind to a star as the natal cluster of stars dissolves \citep{2012Perets}, at an efficiency ranging from 1\% to 9\%. We wondered whether a new mechanism, in which the planet stays with its parent star, could populate the same region. 

\subsection{Analogy to the Formation of Cometary Orbits} \label{overview}

The formation mechanism we study here is closely analogous to the leading mechanism thought to populate the Solar System's outer reaches with comets. \cite{1950Oort} realized that bodies could transition to very distant orbits by first scattering off the known planets, which raises their aphelion to thousands of AU while the perihelion remains among the planets, and in such a state the passage of stars in the Galaxy can raise the perihelion up out of the planetary region. Then the objects would be put in ``cold storage'' and occasionally return later, warming up and sprouting a tail as a comet.

Simulations of the formation of the Solar System's Oort cloud have shown a relatively inefficient process, beginning with an initial rapid population by Jupiter and Saturn and followed by a more gradual period of growth mostly due to Uranus and Neptune, where the fraction of bodies making it to the Oort cloud ranges between 5\% and 7.6\% \citep{2004Dones}.

\cite{1993Tremaine} extended this idea to planetary perturbers other than the Solar System's giant planets, inspired by the first discoveries of exoplanets around a pulsar. He leveraged a different external torquing mechanism, that of the Galactic tidal field \citep{1986Heisler}, which makes the problem especially well-posed. He found that the slower scattering of planetesimals off Neptune-mass planets most efficiently populates the clouds around other stars (called exo-Oort clouds by \citealt{2014Veras,2015Stone}).

The most comprehensive model of the Solar System Oort cloud, including the effects of an early cluster environment, has been made by \cite{2006Brasser}. They calculated how objects that are being scattered out by planets can be saved from ejection by passing stars. This Oort mechanism can produce objects on orbits similar to Solar System objects such as Sedna (e.g. \citealt{2004Morbidelli}, mechanism 4). In this paper we follow some of the assumptions and methods of the simpler and earlier model of \cite{2000Fernandez}.

A population of high-apastron planets would be needed in order to emplace a wide-orbit planet via this mechanism. Planet-planet scattering has been shown to explain most of the eccentric gas giant planets (e.g., \citealt{2008Juric,2008Chatterjee}). This would give a source of planets on their way to be ejected, that can become stranded in the Oort cloud by passing stars. \cite{2009Veras} have computed the distribution of long-period planets as a function of time using this concept, and they find $>$10\% of systems undergoing scattering have an outermost planet beyond 100 AU at a time of 5 Myr. This population of planets then decreases over time as the planets are ejected from the system; we propose instead a mechanism to save these planets from ejection and maintain a wide orbit for Gyr timescales. \cite{2009Veras} mention the possibility of giant planets being perturbed onto an Oort-like orbit, but due to an expected low efficiency as well as the difficulties in observing such old and cold giant planets via direct imaging, their work focuses on young giant planets in the process of being scattered and eventually ejected.

\subsection{Cluster environment}

By comparing the counts of young clusters and the rate at which stars in the field of the galaxy were formed, \cite{2003Lada} concluded ``embedded clusters may account for a large fraction of all star formation occurring locally.''  If these clusters survive the embedded phase, close encounters with other stars are required to cause individual stars to evaporate from the cluster and reach the field of the galaxy. Usually, it is around stars in the field of the Galaxy, not in star clusters, that exoplanets are found. However, some clusters may dynamically disrupt via prompt evaporation of the gas owing to photoevaporation by the most massive stars, which would make the cluster phase brief and cut down the number of encounters. Some investigators recently found that \textit{Gaia} parallaxes show the Scorpius-Centaurus OB association was never highly clustered \citep{2018Wright,2018Ward}. Indeed, statistical clustering of star-forming regions need not imply gravitational binding \citep{2018Elmegreen}. Nevertheless, the main paradigm remains that most field stars, and hence most exoplanet hosts, formed in clusters and therefore endured flybys of other stars early in the system history \citep{2010Adams}. 

\subsection{Plan for this paper}

We describe numerical methods by which we explore the mechanism in section~\ref{sec:methods}. In section~\ref{sec:results} we show examples of the mechanism, discuss its dependence on planetary and cluster parameters, and estimate its efficiency. Finally, in section~\ref{sec:conc&disc}, we discuss the relation to other mechanisms and propose observational tests of the theory. 

\section{ Methods } \label{sec:methods}

Our calculations follow those in the planet-planet scattering literature -- a distribution of planets will begin scattering and place some on long-period orbits. Our treatment of what happens next diverges from the literature. Instead of placing an outer boundary (500 AU is common), we will allow planets to remain in the simulation and simulate passing stars. The effect of these flybys could raise the scattering planet's periastron through torques, saving it from further scattering, an effect which would have been missed in previous calculations. Contrariwise, the flyby could tidally pull the planet away, in which case it would be a true ejection, taken away from the planetary system. 

\subsection{ Initial Conditions } \label{IC}

The distributions of initial conditions were chosen in a somewhat arbitrary fashion to provide a wide array of potential planet-planet scattering encounters. The goal is not to have a random sampling of systems as close to the distribution in nature as possible; the goal is to produce a few examples of models in which the planet-saving mechanism occurs, to learn how it operates. The starting point for the simulations was assumed to be after the dissipation of the gaseous protoplanetary disk, once gas giant planets have accreted their massive envelopes.

\begin{figure}[ht]
\centering
\includegraphics[width=1\linewidth]{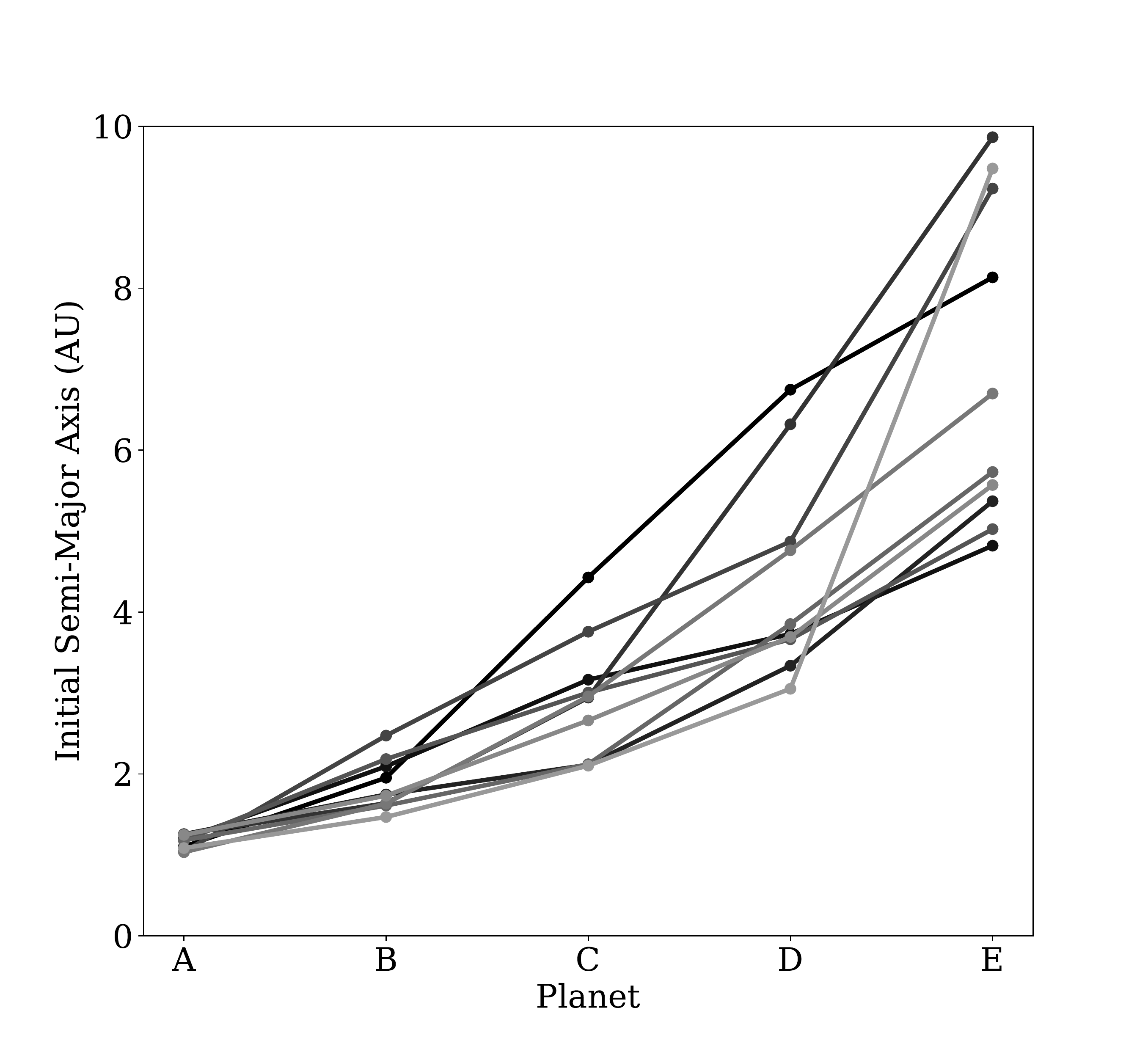}
\caption{Initial semi-major axes over ten different simulations. Planets from the same simulation are shown in the same shading.}
\label{fig:initial_SMA}
\end{figure}

Planetary masses were chosen uniformly in log $M$ between $M$ = 0.1 and 10 $M_{J}$. The semi-major axis for the innermost planet was chosen uniformly in log $a$ between log $a$ [AU] = 0.0 and 0.1. The remaining semi-major axes were assigned at approximately 5 mutual Hill radii separation \citep{1996Chambers}. If the outermost semi-major axis was beyond 10 AU, a new distribution of masses was drawn and the semi-major axis assignment repeated. A distribution of initial semi-major axes is shown in Figure~\ref{fig:initial_SMA}.

The inclination and eccentricity for each planet were drawn from a Rayleigh distribution with scale parameters of 0.1 radians and 0.01, respectively. The ascending node, argument of pericenter, and mean anomaly were all chosen uniformly between 0 and $2\pi$ radians. 

The radius of each planet was assigned using the mass-radius prediction function from the \texttt{Forecaster} package \citep{2017Chen}, which forecasts a radius using a random draw from a probabilistic mass-radius relation. Both the central star and flyby star were taken to have solar mass, and the central star solar radius. Planet-planet and planet-central star collisions were allowed during the simulation.

\subsection{Simulation Methods} \label{simmethods}

The simulations were computed using the \texttt{REBOUND} package \citep{2012Rein} with initial conditions as described in section~\ref{IC}. Computations were performed using the IAS15 integrator \citep{2015Rein}, a 15th-order adaptive-step integrator that can handle close encounters and highly eccentric orbits, checking for ejections (distance $>$2 x 10$^{5}$ AU) every 0.5 years. Positions, velocities, and astrocentric orbital parameters were recorded every 5 years through the duration of the simulation. Initial runs were computed over 1 Myr while allowing for flyby stars.

Flyby stars were inserted into the simulation in the event that the planets were scattering, as determined by having a planet with a semi-major axis of greater than 20 AU. Each star's initial location was randomly placed on a 1000 AU sphere centered on the central star with an inward-pointed velocity vector randomly drawn with a one-dimensional dispersion of 1 km s$^{-1}$ and probabilistically weighted with the cosine of the incidence angle via rejection sampling \citep{1972Henon}. The flyby star was removed when it again passed 1000 AU, and another star was sent flying by (if the scattering criterion, any planet's $a > 20$ AU, was met) after a delay of 10 kyr, a timescale that allows any violent instabilities in the system to resolve.

There were ten separate initial runs, which were the parent simulations. The subset of each parent simulation during the passage of a flyby star is a child simulation, which can be considered as independent events due to the time delay between insertions. There were 230 total child simulations. The planetary properties of mass and radius were the same for all the child simulations of a given parent simulation; however, the orbital properties of the planets changed for each child simulation due to the interactions occurring between flyby star passages.

After completion of the initial parent simulation, the child simulation results were examined for potential stranded wide-orbit planets due to a stellar passage. If such a result was seen, a new simulation was initiated from the point of the star's removal and computed without further flybys over longer timescales (on the order of 10-100 Myr), recording every 50 years, to determine the stability of the post-flyby configuration of the system, particularly looking at the variability of the semi-major axis of the Oort planet.

\section{Results} \label{sec:results}

\subsection{Example Simulation} \label{ex_sim}

\begin{figure*}[ht]
\centering
\includegraphics[width=1\linewidth]{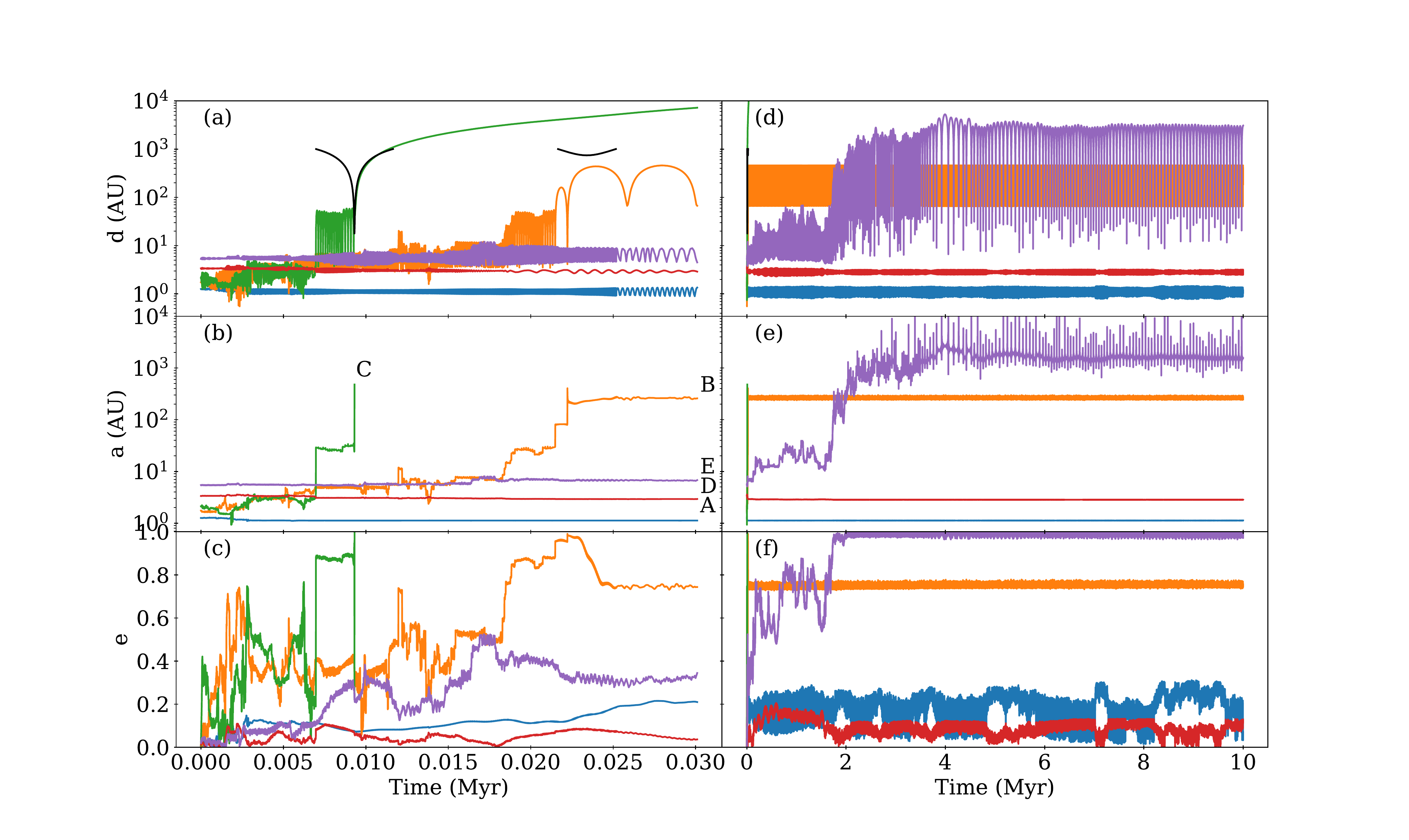}
\caption{The distances (relative to the host star) of the planets (colored) and flyby stars (black), and the semi-major axes and eccentricities of the planets over time. Shown on the left (a-c) is the first 30 kyr of the system with flyby stars. Shown on the right (d-f) is the system over 10 Myr.}
\label{fig:run9b3plots-10Myr}
\end{figure*}

Initial conditions for one example of the simulations described in section~\ref{simmethods} are shown in Table~\ref{run9IC} and the results shown in Figure~\ref{fig:run9b3plots-10Myr}. The left panels (\ref{fig:run9b3plots-10Myr}a-c) show the first 30 kyr of the initial run, where the inserted stars are shown in black in \ref{fig:run9b3plots-10Myr}a. While the first flyby star (undeflected $b = 263$~AU) has a deep, strongly gravitationally focused encounter that ejects planet 9C (green) from the system entirely, the second flyby star (undeflected $b = 827$~AU) appears to lift the periastron of the scattering planet 9B (orange) sufficiently to prevent further scattering and leave the planet on a stable orbit at a semi-major axis of approximately 300 AU. The right panels (\ref{fig:run9b3plots-10Myr}d-f) show the addition of the second simulation over 10 Myr of the system. Planet 9B appears to be stable on its wide orbit resulting from the stellar passage.

After the stellar passages, an additional planet, 9E (purple) undergoes significant orbital excitation from the inner planets. It would be ejected on a shorter timescale than observed, were it not for the intervention of the now-Oort planet, 9B, which lifts its periastron at around 2 Myr in the simulation. The two planets apparently enter a stage of quasi-stability. Planet 9E has an orbital energy closer to zero, so any strong encounters between it and 9B will tend to eject 9E, leaving 9B as an Oort planet. 

This simulation run is included as an illustration of the mechanism described in section~\ref{overview}. Similar results were seen in other simulations.

\begin{table*}[ht]
\centering
\caption{Initial Conditions for Sample Simulation (run9)}
\resizebox{\textwidth}{!}{%
\begin{tabular}{ccccccccc}
\hline
\hline
Planet Name & Semi-Major Axis & Mass          & Radius        & Eccentricity & Inclination & Ascending Node & Argument of Pericenter & Mean Anomaly \\
            & $AU$            & $M_{Jupiter}$ & $R_{Jupiter}$ &              & $deg$       & $deg$          & $deg$                  & $deg$        \\
            \hline
9A          & 1.2563          & 0.6544        & 1.2229        & .0045        & 5.2181      & 9.1809         & 280.4906               & 171.6866     \\
9B          & 1.7460          & 0.1668        & 0.9838        & .0054        & 7.9660      & 229.1612       & 43.2556                & 70.8627      \\
9C          & 2.1131          & 0.1298        & 0.4769        & .0083        & 9.7999      & 220.6738       & 310.6611               & 102.6931     \\
9D          & 3.3368          & 2.1931        & 1.0545        & .0145        & 0.8957      & 239.1286       & 39.4472                & 176.9094     \\
9E          & 5.3686          & 0.1458        & 1.3875        & .0153        & 5.6120      & 238.2865       & 186.4250               & 32.6289     \\
\hline
\end{tabular}%
}
\label{run9IC}
\end{table*}

\subsection{Efficiency}

The expected efficiency of saving a planet on a wide orbit of $a$ is given by
\begin{equation}\label{fsaved}
    f_{saved,tot}(a)= N(b_{min}<10a)p(a)f_{strong}f_{save},
\end{equation}
where $N(b_{min}<10a)$ is the expected number of relevant stellar passages with an impact parameter within $10a$, $p(a)$ is the probability of a planet being at a given semi-major axis, $f_{strong}$ is the fraction of strong encounters for $b/a<10$, and $f_{save}$ is the fraction of strong encounters that result in a save. Each of these components will be defined and discussed in detail in the following sections.

\subsubsection{Stellar Passage Frequency} \label{stellar_passage_freq}

In order to understand the feasibility and efficiency of this method of producing stable, wide-orbit planets, the frequency of close stellar passages must be evaluated.

Following the lead of \cite{2000Fernandez}, the natal star cluster is taken to have an initial number density of 25 stars pc$^{-3}$ for stars of approximately 1 $M_{\odot}$ and characteristic relative velocities of 1 km s$^{-1}$. The cluster is assumed to dissipate over a timescale of 10$^{8}$ years with the density decreasing linearly to zero over the lifetime of the cluster. This assumption neglects the effects of gas in the cluster. Having gas in the cluster for longer reduces the dissipation rate of the cluster and increases the number of close stellar encounters \citep{2009Proszkow}, making this a conservative assumption. Further, the average stellar mass in nearby clusters is 0.5 $M_{\odot}$ \citep{2009Proszkow}, so our assumption that all the stars are 1 $M_{\odot}$ reduces the number of stars in the cluster and thereby reduces the interaction rate.

In addition to the relatively ``dense'' cluster (actually less dense than many stellar clusters), a ``loose'' cluster with initial density 10 stars pc$^{-3}$ was also considered, with the same dissipation profile.

The rate of flyby stars within a specified impact parameter, $b_{max}$, is given by 
\begin{equation}\label{pass_rate}
    \Gamma = n\pi b_{max}^{2}v,
\end{equation}
where $n$ is the number density of stars in the cluster and $v$ is the incoming relative velocity. The effect of gravitational focusing is neglected. For $v_\infty$ = 1 km s$^{-1}$, this approximation is appropriate beyond $\sim$1000 AU.

\begin{figure}[ht]
\centering
\includegraphics[width=1\linewidth]{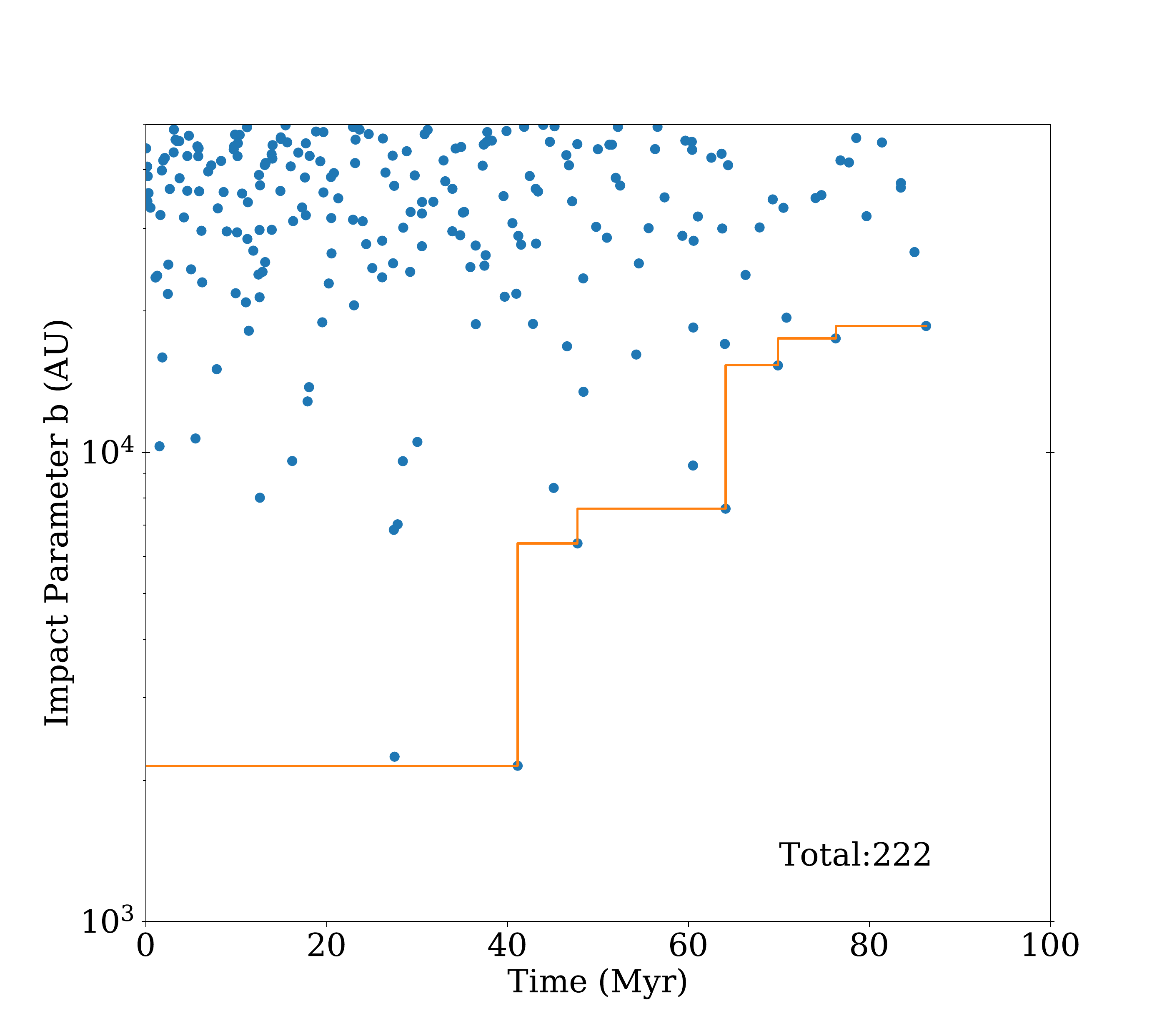}
\caption{The impact parameter of passing stars shown over the life of one sample cluster with an initial density of 25 stars pc$^{-3}$. The orange line connects the recent closest passages, i.e., those without any future closer passages.}
\label{fig:flybys_over_time}
\end{figure}

For statistical modeling, we calculate the passages with an impact parameter of less than 5 x 10$^{4}$ AU over the life of the cluster. A sample draw is a choice of $(b/b_{max})^2$ from a uniform distribution between 0 and 1. The time between flybys was modeled as a Poisson process at the rate specified in equation \ref{pass_rate}. Because the density is not constant over time, the wait time between passages was corrected to account for the change in density over the uncorrected wait time (see Appendix for details). The results of one example cluster model simulation are shown in Figure~\ref{fig:flybys_over_time}. Close passages can disrupt the planetary system (as demonstrated in Figure~\ref{fig:run9b3plots-10Myr} with the ejection of planet 9C) in addition to ``saving'' the planets, so the relevant passages that we consider in our analysis are the recent closest passages --- that is, the closest passages \emph{without any future closer passages}.

\begin{figure}[ht]
\centering
\includegraphics[width=1\linewidth]{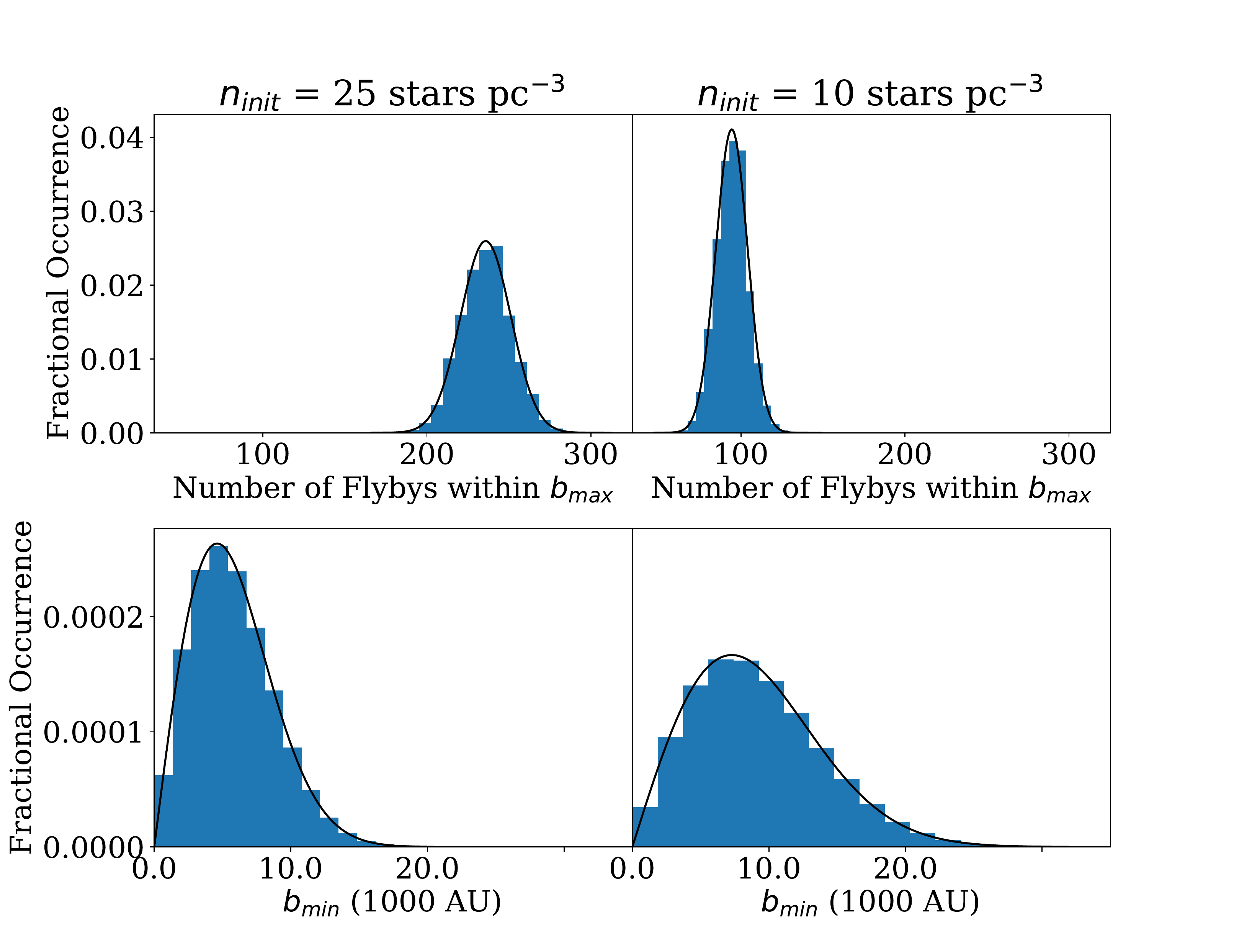}
\caption{Distribution of the total number of flybys (top) and the closest impact parameter (bottom) from 10$^{6}$ model simulations. The line in each panel shows the probability distribution function. The left panels show the distribution for dense clusters (initial density of 25 stars pc$^{-3}$) with loose clusters (initial density of 10 stars pc$^{-3}$) on the right.}
\label{fig:25_10_cluster_dists}
\end{figure}

A distribution of the total number of passages and the single closest impact parameter over 10$^{6}$ simulations is shown for both a dense cluster (25 stars pc$^{-3}$) and a loose cluster (10 stars pc$^{-3}$) in Figure~\ref{fig:25_10_cluster_dists}. The probability distribution for the total number of flybys is Poisson, which approaches a normal distribution with mean $N=\int \Gamma dt$ and standard deviation $\sqrt{N}$. For the closest impact parameter, $b_{min}$, the probability distribution function is given by 
\begin{equation}\label{bmin_pdf}
    p(b_{min})=\frac{2Nb_{min}}{b_{max}^{2}}(1-\frac{b_{min}^{2}}{b_{max}^{2}})^{N-1}.
\end{equation}

In the limit as $N \rightarrow \infty$ and $b_{max}\rightarrow \infty$, $b_{min}$ is distributed as a Rayleigh distribution with scale parameter $\sigma = \sqrt{\frac{b_{max}^{2}}{2N}}$, which is plotted in Figure~\ref{fig:25_10_cluster_dists}, bottom panels. See Appendix for details of derivation. 

Both distributions are dependent on the density of the cluster. Analytically, the total number of passages within the specified $b_{max}$ is expected to increase linearly with the initial density of the cluster. The median closest passage is expected to decrease as the square root of the initial density of the cluster. This was consistent with our simulations at varying densities.

\begin{figure}[ht]
\centering
\includegraphics[width=1.0\linewidth]{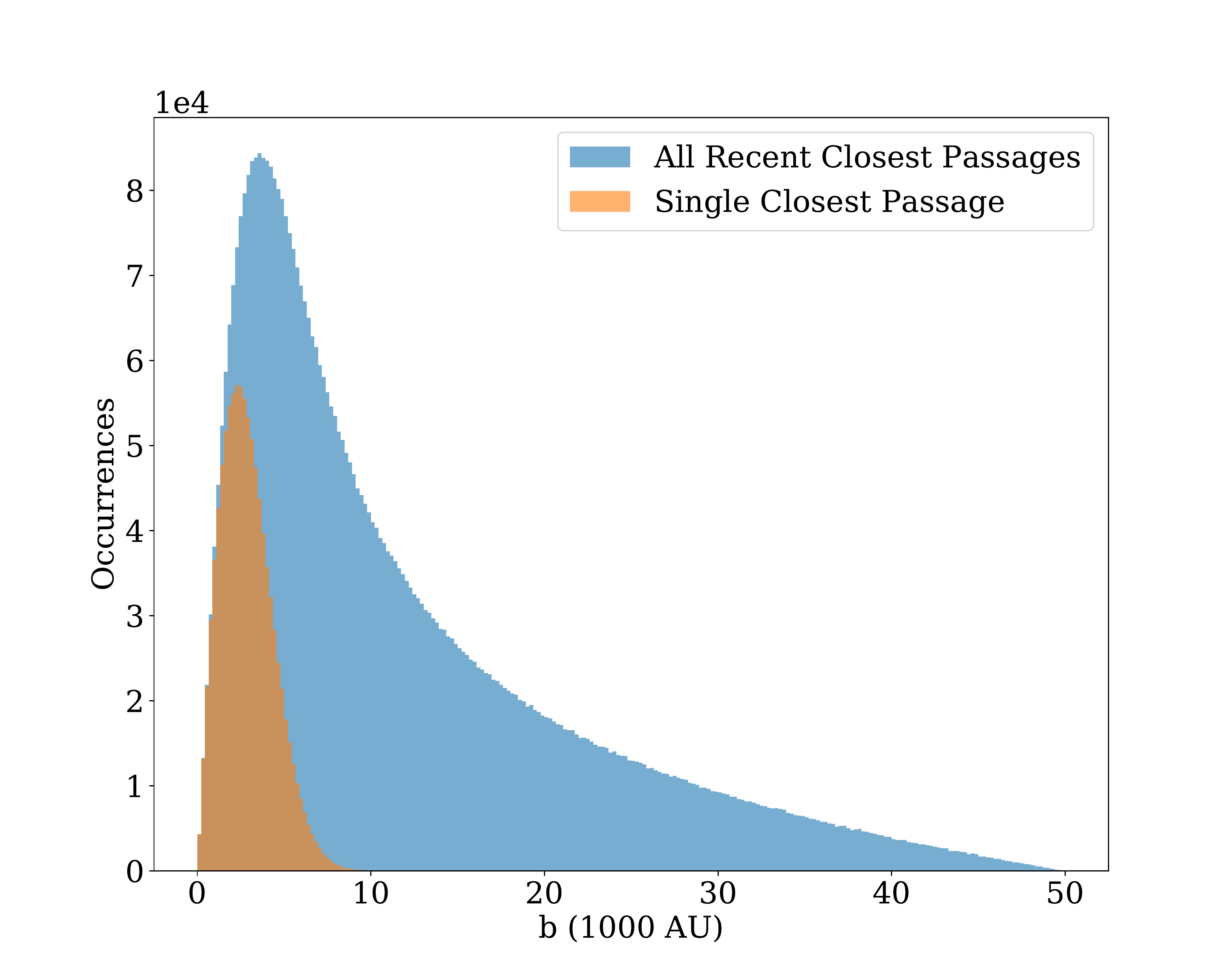}
\caption{The distribution of all recent closest passages over the lifetime of the cluster over 1 million simulations, compared to the distribution of only the closest passage of each simulation.}
\label{fig:all_bmin}
\end{figure}

The $b_{min}$ distribution in Figure~\ref{fig:25_10_cluster_dists} is for the single closest passage during the cluster's lifetime. Including all recent closest passages, as illustrated in Figure~\ref{fig:flybys_over_time}, increases the number of flybys. The average cluster, with an initial density of 25 stars pc$^{-3}$, has 5 of these recent closest flybys. Figure~\ref{fig:all_bmin} shows how the $b_{min}$ distribution differs for the nominal cluster by including all recent closest flybys. While the single closest passage distribution is Rayleigh-like, the distribution for all recent closest passages has a fatter tail and increased occurrences of more distant flybys.

While our statistical approximation is a simplified approach, it is consistent with more detailed stellar cluster simulations. Using dynamical N-body simulations, \cite{2009Proszkow} find that, depending on cluster parameters, the total rate of close stellar encounters less than 5 x 10$^{4}$ AU ranges between 0.5 and 13.5 Myr$^{-1}$ in the first 10 Myr (assuming our 25 stars pc$^{-3}$ are in a sphere of radius $\sim$1 pc and therefore N $\approx$ 100). Our approximation gives an average of 4.5 Myr$^{-1}$ encounters in the first 10 Myr.

\subsubsection{Stellar Passage Impact} \label{stellar_passage_imp}

\begin{figure}[ht]
\centering
\includegraphics[width=1\linewidth]{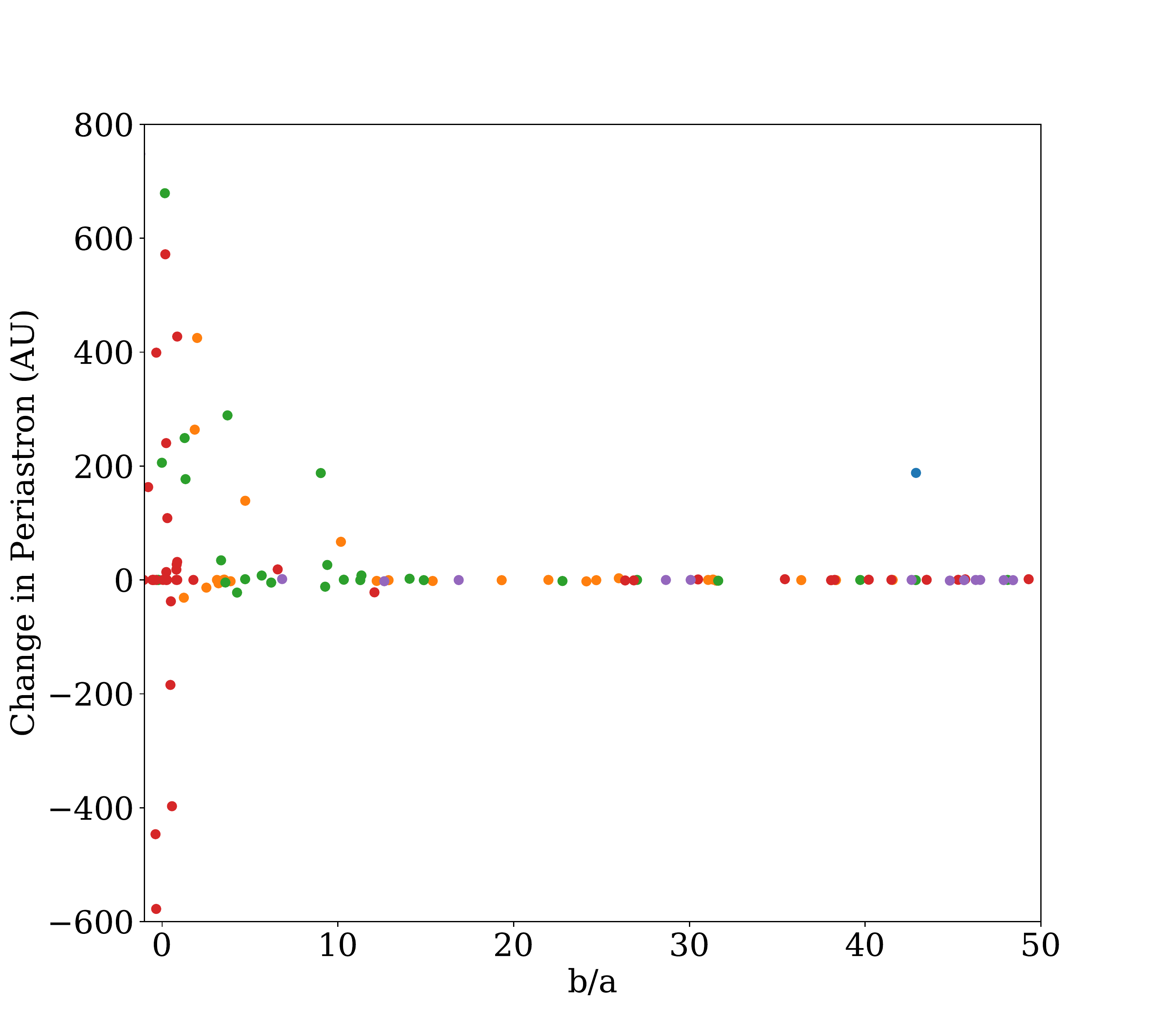}
\caption{The change in a planet's periastron as a function of the ratio of the flyby star's impact parameter to the planet's semi-major axis. Colors correspond to different lettered planets in each system with the same initial ordering as Figure~\ref{fig:run9b3plots-10Myr}a, e.g., blue points are planet `A's.}
\label{fig:b2a}
\end{figure}

A flyby star is most likely to have a strong impact on planets that have a semi-major axis of the same order of magnitude as the flyby's impact parameter. As can be seen in Figure~\ref{fig:b2a}, with only a couple exceptions, the vast majority of encounters resulting in a significant change in periastron of the planet happened for $b/a \lesssim 10$. 

Examining these encounters, we can determine the fraction of encounters with $b/a < 10$ that result in a strong encounter. A strong encounter is defined to be one where the planet's periastron changes by more than 20 AU. We find that there were 51 encounters with $b/a < 10$ and 20 of these were strong encounters, giving an $f_{strong}$ of 0.39 $\pm$ 0.07.

For strong encounters seen during our simulations, we find that there are five likely outcomes. The planet may have its periastron lifted (generally resulting in being ``saved''); an already unbound planet from scattering may become rebound to the host star (also a save, although more dependent on specific star-planet interaction); the planet may become unbound, either due to planet-planet scattering or from an interaction with the flyby star; the planet may have its periastron reduced (potentially putting a previously safe planet into danger of scattering); or the change may be due to planetary interactions where the flyby star appears to have no effect.

Across the simulations performed, there were 230 flyby stars. Of these 230, 30 resulted in a strong encounter. These 30 encounters are categorized as described above, resulting in the branching probabilities shown in Table~\ref{outcomes}. Combining the two outcomes that save the planet, the simulated $f_{save}$ is 0.27 $\pm$ 0.08.

The uncertainty for $f_{strong}$ and $f_{save}$ is calculated via the normal approximation of the distribution of error for a binomially-distributed observation.

\begin{table}[ht]
\centering
\caption{Rate of Outcomes for Strong Encounters}
\label{outcomes}
\begin{tabular}{lc}
\textbf{Outcome}   & \textbf{Occurrence} \\
Lifts Periastron   & 20\%                \\
Rebinds            & 7\%                 \\
Unbound            & 43\%                \\
Reduces Periastron & 17\%                \\
No Flyby Effect    & 13\%                
\end{tabular}
\end{table}

\subsubsection{Distribution of Planets During Scattering} \label{dist_sma}

In order for a planet to be saved on a wide orbit by a flyby star, there must first be a planet that has been scattered out to a large semi-major axis at the time that the flyby star passes. The expected distribution of semi-major axes over time depends on the properties of the scattering planet; it must be massive enough to scatter another object to a certain distance in the given system age. \cite{1993Tremaine} found that a necessary condition on the interior planet doing the scattering for the formation of an Oort-type comet cloud is

$$\frac{M_P}{M_\oplus } \gtrsim \left( \frac{M_*}{M_\odot} \right)^{3/4} \left( \frac{t_*}{10^9\, yr} \right)^{-1/2} \left( \frac{a_P}{1\, AU} \right)^{3/4}.$$

For our range of planet parameters (0.1 $\leq$ $M_P/M_J$ $\leq$ 10; 1 $\leq$ $a_P$/1 AU $\leq$ 10), this condition is met within the cluster's lifetime ($t_*$ = $10^8$ yrs). Considering in our case the scattered bodies are planets and not comets, the scattering planet must also be at least as massive as the scattered planets \citep{2008Ford}.

To derive the distribution of semi-major axes of scattered planets, an N-body simulation was run having one planet with parameters chosen in a similar manner as described in section~\ref{IC} (with N = 1). The semi-major axis of the planet was chosen uniformly in $a$ between 1 and 10 AU. The randomly generated planet had $M$ = 0.59 $M_J$ and $a$ = 9.49 AU. 100 test bodies were added with initial semi-major axes drawn uniformly within 10\% of the planet's semi-major axis and other orbital parameters drawn in the same manner as the planet's. The test bodies were massless in order to decrease computation time. The effect of including the mass of the planets being scattered would change the outcome of any given energy exchange interaction by a factor of $\sim$2, which would not appreciably alter the resulting distribution. The system was integrated for $10^8$ years.

In order to examine the dependence of the distribution on the mass of the scattering planet, this simulation was repeated with planet masses at the limits of the range, $M$ = 0.1 $M_J$ and $M$ = 10 $M_J$. The same semi-major axis ($a$ = 9.49 AU) was used for consistency.

\begin{figure}[ht]
\centering
\includegraphics[width=1\linewidth]{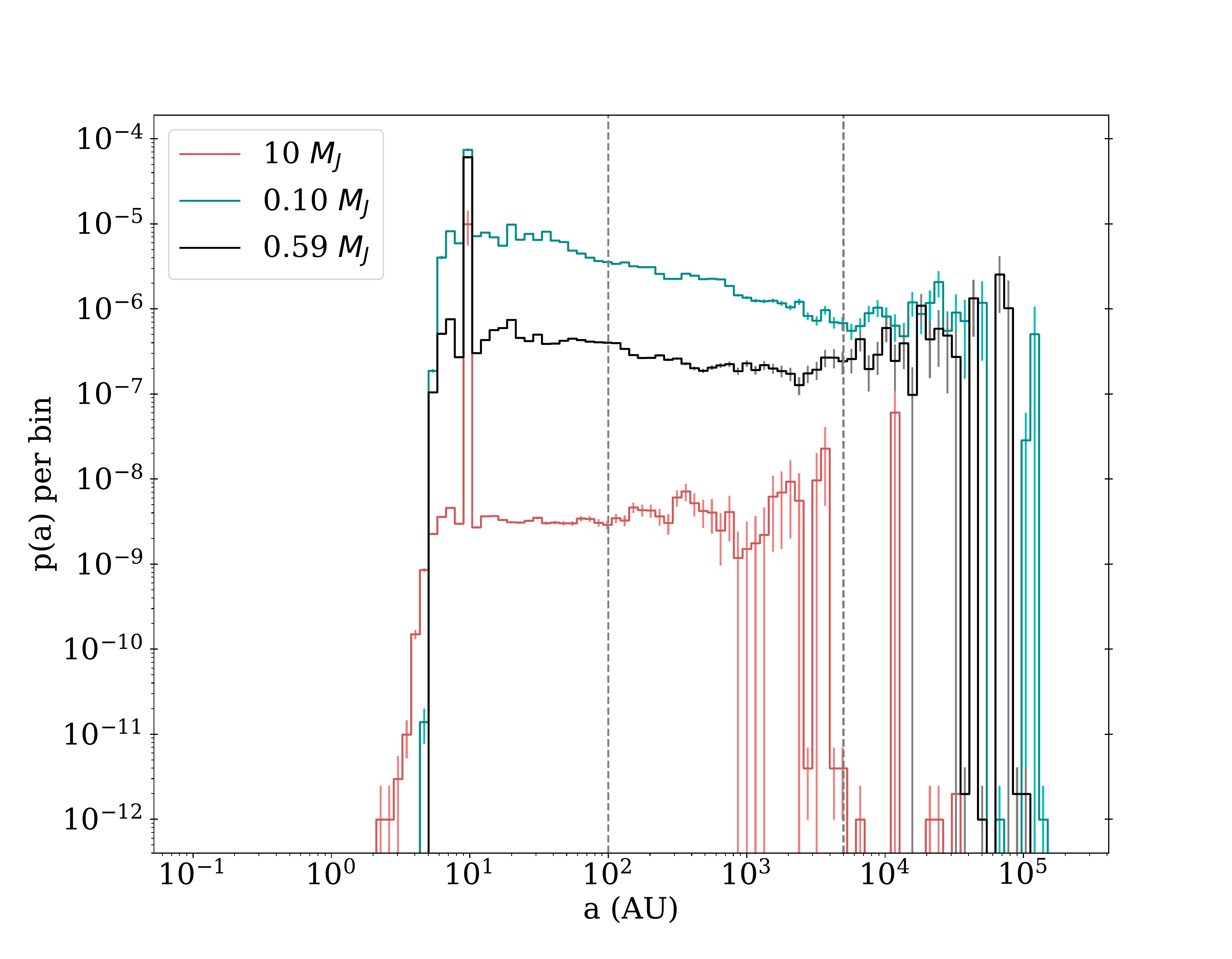}
\caption{The probability of finding a body at a given semi-major axis. The probability is shown over 100 bins of equal log-space width between $10^{-1}$ and $10^{5.3}$ AU. The vertical dashed lines highlight our area of interest between 100 and 5000 AU. The calculated $p(a)$ distributions are shown for 3 different scattering planet masses at the same scattering planet semi-major axis. The peak at $\sim$10 AU is due to the location of the scattering planet and the initial positions of the bodies.}
\label{fig:scattering}
\end{figure}

Taking the amount of time spent at any given semi-major axis (at a resolution of 20 years) and averaging over the lifetime of the cluster (assumed to be $10^8$ years) and the number of bodies gives a probability for any one body to be at a given semi-major axis at any point in cluster's life. The resulting $p(a)$ distribution is shown in Figure~\ref{fig:scattering}. At the range of semi-major axes that we are interested in, the probability is relatively flat, dropping less than 1 dex from 100 to 5000 AU for the $M$~=~0.59~$M_J$ scatterer. This is due to the fact that bodies that are scattered out to a wide semi-major axis will spend a longer time there before being scattered again, as their period becomes very large and therefore the number of interactions with the scattering planet decreases. For the larger mass scattering planet, the distribution becomes flatter but also reduces in amplitude, as bodies are ejected much more quickly from the system. In addition, the maximum semi-major axis reached before ejection moves inward, seen in the sharp decline in the $p(a)$ distribution. The lower mass scatterer keeps bodies in the system for much longer, giving a higher overall amplitude, but is slower to move the bodies to high semi-major axes, resulting in a steeper probability distribution and a decline in the distribution at larger semi-major axis.

\subsubsection{Time Dependence}

The passage of time in the cluster affects the circumstances related to this mechanism in two ways. First, planets are likely to begin scattering early in the lifetime of the cluster. As time increases, more planets are ejected until the system reaches a stable configuration, so the probability of having a planet at a large semi-major axis decreases. Secondly, as the cluster dissipates and the stellar density decreases, the single closest flybys become less frequent and the recent closest flybys become more frequent but with a much larger impact parameter.

\begin{figure}[ht]
\centering
\includegraphics[width=1\linewidth]{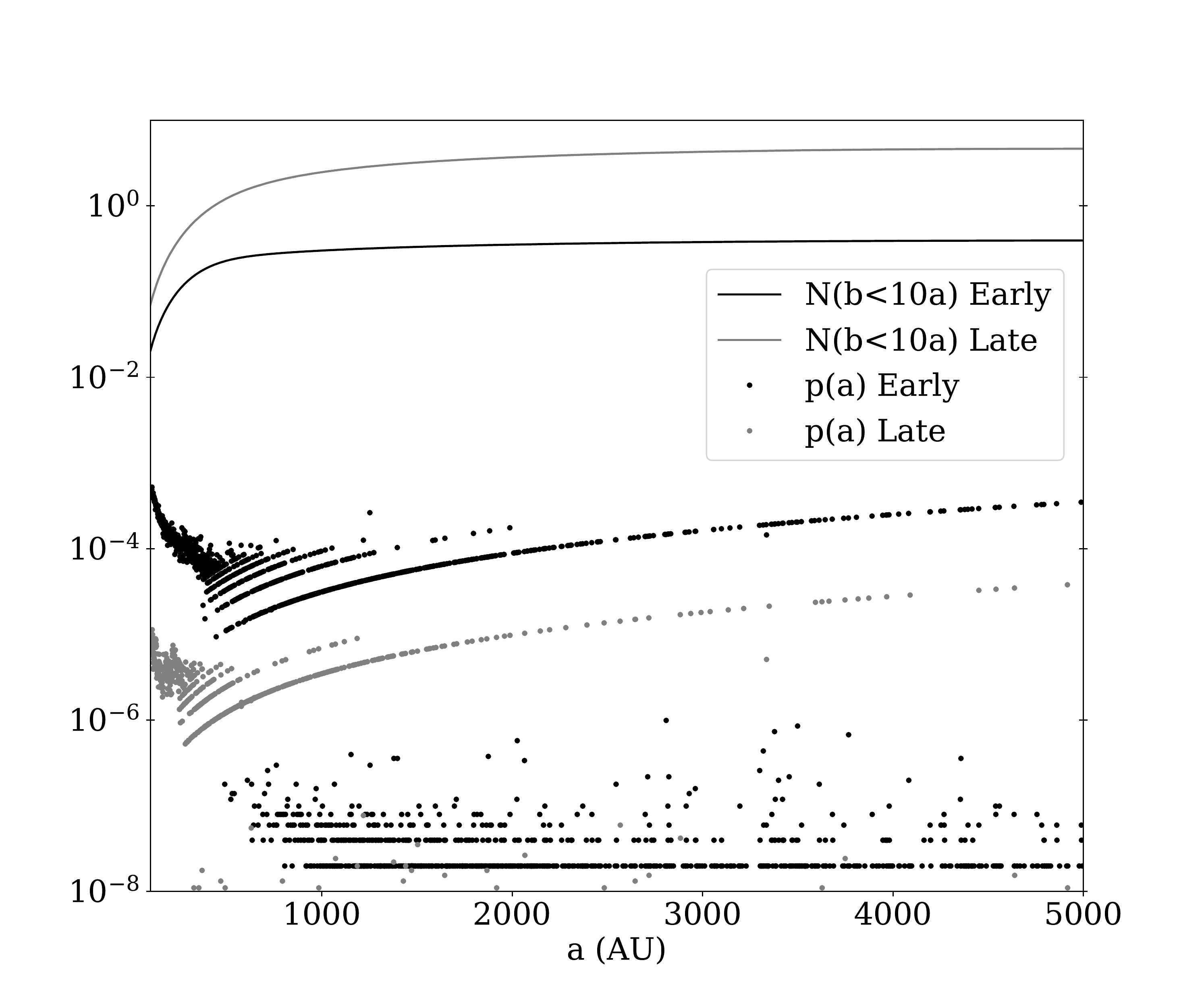}
\caption{The $N(b_{min}<10a)$ and $p(a)$ (for $M_{scattering}$ = 0.59 $M_J$) factors (from equation \ref{fsaved}) over two time periods of the cluster. The $p(a)$ is a computed with bin width of 1 AU.}
\label{fig:time_dependence}
\end{figure}

By splitting the time of the cluster into two time periods, an early period from 0 to $10^7$ years and a late period from $10^7$ to $10^8$ years, we can capture some of this time dependence. The effects of the time period splitting on both the number and impact parameter of the flyby stars and the probability of having a planet at a given semi-major axis are shown in Figure~\ref{fig:time_dependence} for the $M$~=~0.59 $M_J$ scatterer. The $p(a)$ is the same as the black curve in Figure~\ref{fig:scattering}, except here the spacing is 1 AU and shown linearly in semi-major axis and split between the two time periods. As expected, the probability of having a planet at a large semi-major axis decreases significantly in the late period. The $f_{strong}$ and $f_{save}$ factors are unaffected by the timing of the flyby event.

The single closest flyby, shown in the bottom panels of Figure~\ref{fig:25_10_cluster_dists}, has a probability per unit time which is almost twice as much in the early period compared with the average of the late period. As seen in Figure~\ref{fig:all_bmin}, these closest flybys usually have a much lower impact parameter. This explains why the difference between $N(b_{min}<10a)$ in the early and late periods is much less at the lower end of the semi-major axis space. However, because our area of interest is at large semi-major axes, even the higher impact parameters are within a range to have a potential effect ($b/a < 10$), and so the fact that the recent closest flybys are more frequent in the late period is the dominant effect. This can be seen in Figure~\ref{fig:time_dependence} most strongly for planets with $a$ = 5000 AU, where on average there are fewer than 1 recent closest flybys within 50,000 AU ($b = 10a$) in the first $10^7$ years, but there are $\sim$4 within that value between $10^7$ and $10^8$ years.

\subsubsection{Total Efficiency}

\begin{figure}[ht]
\centering
\includegraphics[width=1\linewidth]{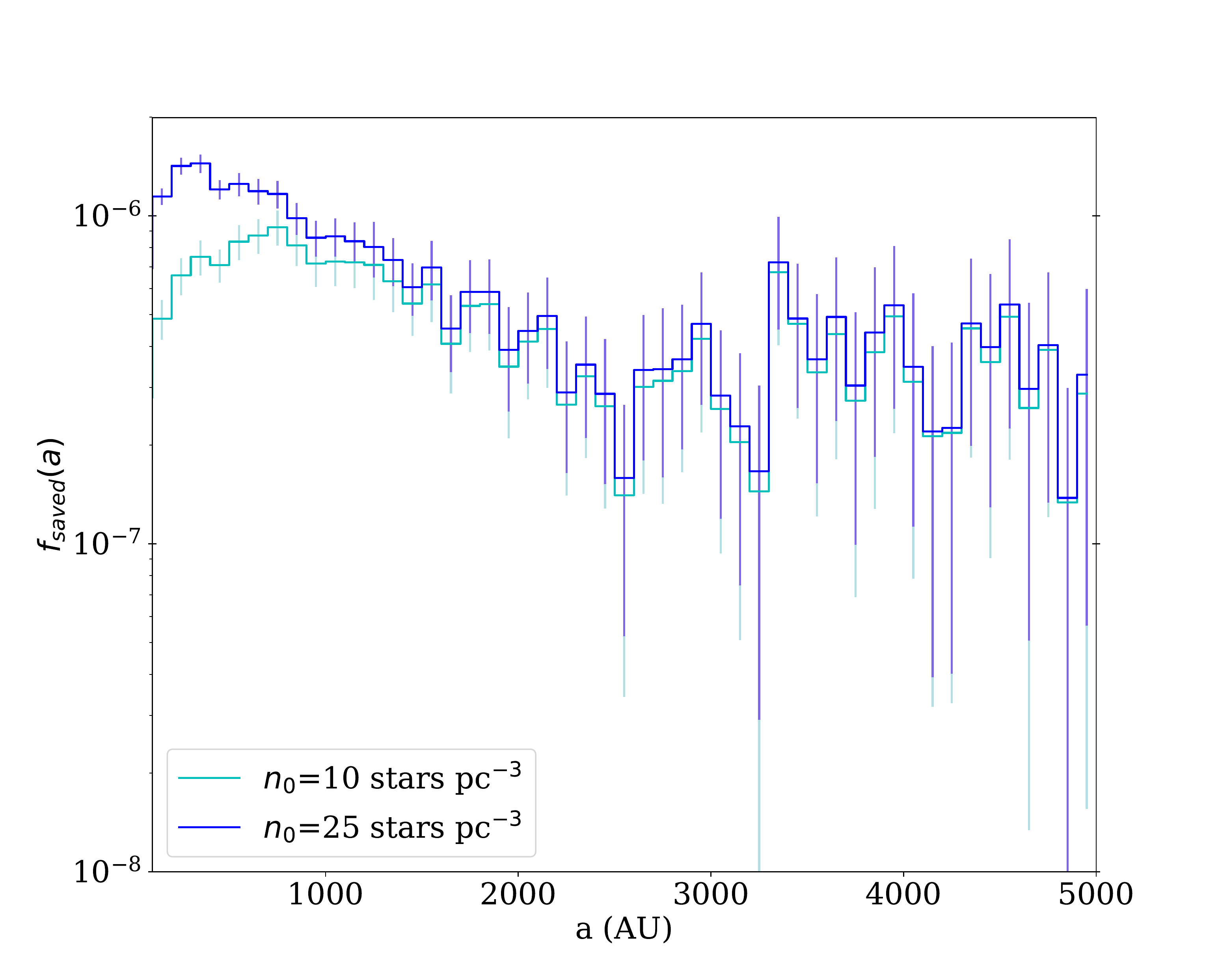}
\caption{The efficiency of saving a planet given the planet's semi-major axis for both the nominal cluster and the loose cluster; bin width is 100 AU for visibility.}
\label{fig:efficiency}
\end{figure}

Returning to equation \ref{fsaved}, we can now calculate an expression for the total efficiency as a function of a planet's semi-major axis. In each time period, summing the simulated distribution of all $b_{min}$ from 0 to $10a$ gives $N(b_{min}<10a)$. The distribution obtained from the scattering simulation with $M_{scattering}$ = 0.59 $M_J$, split into two time periods as shown in Figure~\ref{fig:time_dependence}, gives $p(a)$. Both of these distributions are calculated with a resolution of 1 AU. Combined with $f_{strong}$ and $f_{save}$ from section~\ref{stellar_passage_imp}, we find an efficiency as a function of $a$ as shown in Figure~\ref{fig:efficiency}.

In order to consider the total expected efficiency of wide-orbit planets resulting from stellar flybys, the distribution shown in Figure~\ref{fig:efficiency} is integrated from 100 AU to 5000 AU ($b_{max}$/10). For the nominal cluster, this total efficiency is 0.282$\pm$0.007\%. For the loose cluster, this total efficiency is 0.226$\pm$0.006\%. The uncertainty cited here is solely from statistical effects and does not incorporate uncertainties related to the simplifying assumptions used in our models.

\begin{figure}[ht]
\centering
\includegraphics[width=1\linewidth]{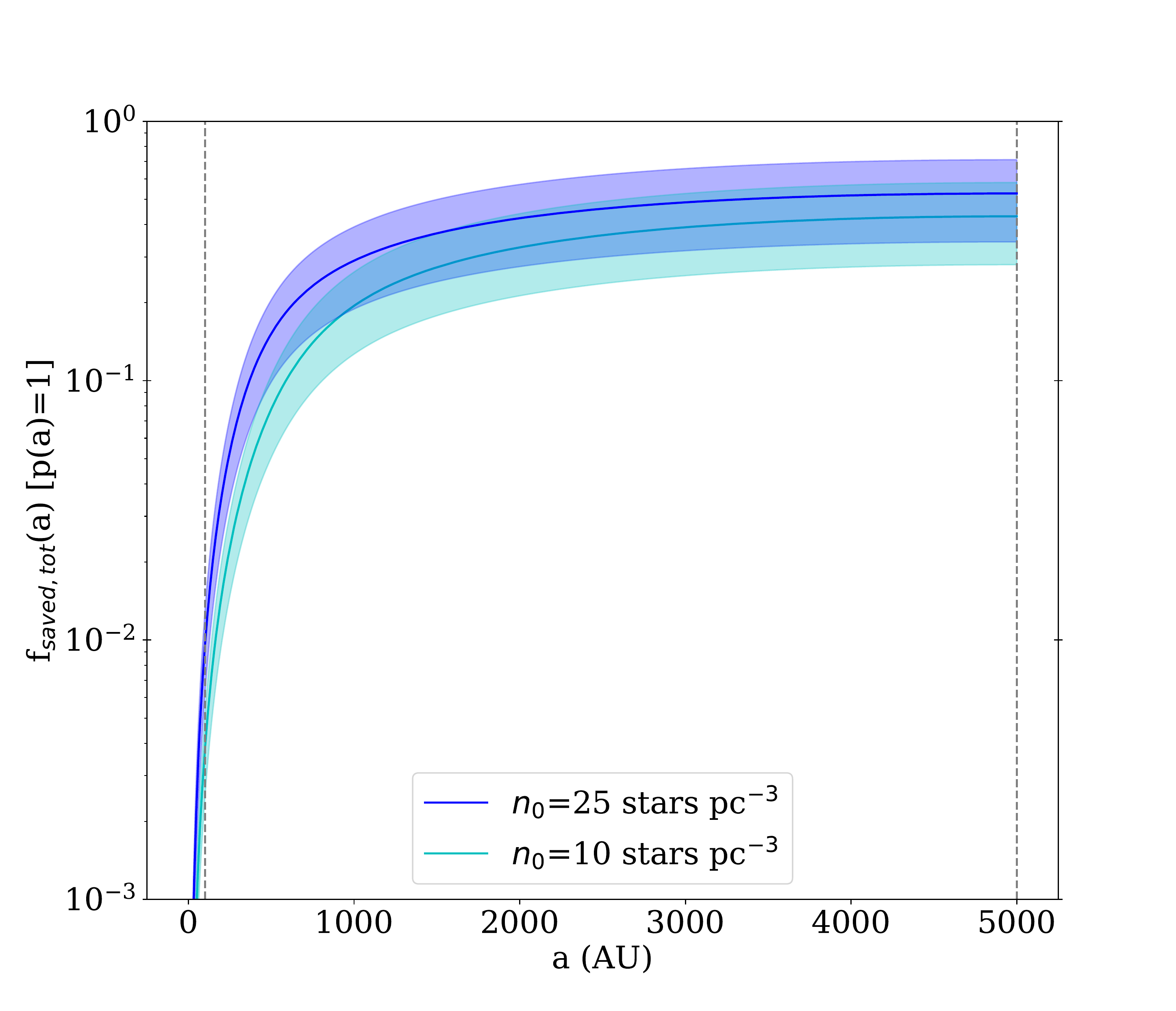}
\caption{The efficiency of saving a planet given the planet's semi-major axis for both the nominal cluster and the loose cluster, assuming a constant $p(a)$ = 1.}
\label{fig:efficiency_nopa}
\end{figure}

The efficiency calculation is strongly dependent on the distribution of semi-major axis used to calculate $p(a)$, as can be seen in Figure~\ref{fig:efficiency_nopa}, which shows the efficiency calculated with the $p(a)$ term set equal to 1. The $p(a)$ distribution itself is sensitive to the properties of the scattering planet, the host star, and the population of planets formed within a system, as well as the stochastic history of planet-planet interactions. This can be seen clearly in Figure~\ref{fig:scattering}, where the variation of mass of the scattering planet has a large effect on the shape and magnitude of the $p(a)$ distribution. 

Therefore the preceding efficiency calculation should be considered a sample calculation given one particular scattering scenario. Repeating the calculation for the other simulated $p(a)$ distributions gives a total efficiency of 0.007$\pm$0.001\% ($M_{scattering}$ = 10 $M_J$) and  1.64$\pm$0.02\% ($M_{scattering}$ = 0.1 $M_J$) for the nominal cluster and 0.006$\pm$0.001\% ($M_{scattering}$ = 10 $M_J$) and 1.07$\pm$0.02\% ($M_{scattering}$ = 0.1 $M_J$) for the loose cluster.

\section{Conclusions and Discussion} \label{sec:conc&disc}

\subsection{Summary of Results}

In this paper, we have explored how planet-planet scattering can be interrupted by passing stars, creating stable wide-orbit planets.  The occurrence of planets between 100 and 5000 AU (Oort planets) by this mechanism is likely to be relatively small; we estimate it is $<$2\% relative to the number of planets that are ejected. The mechanism operates in planetary systems that underwent planet-planet scattering while the host star still orbited in a stellar cluster, which may be the majority of systems that have at least one giant planet. 

The more massive a planet is, the less likely it is to become an Oort planet. It requires another, more massive planet to scatter it out to large distances, and the time spent in the system prior to ejection is smaller for more massive planets. Spending less time at large distances while bound makes massive planets less likely to interact with a close stellar flyby. Loosely-bound Neptune- or super-Earth-mass planets are more likely to become stranded as Oort planets at a rate up to a few percent. 

\subsection{Relation to Other Work}

A completely distinct way a planet might be added on a long-period orbit is by the dissolution of the cluster while free-floating planets are present. The capture probabilities and statistics of such captures have been worked out by \cite{2012Perets} and were seen in cluster simulations of \cite{2019vanElteren}. The semi-major axis distribution of capture-formed planetary systems seen in \cite{2012Perets} (see their Figure 4) peaks at larger semi-major axes than we find for the Oort planets, so these captured wide-orbit planets could form part of a distinct but overlapping population as Oort planets. 

\cite{2011Chatterjee} proposed a method for circularizing eccentric, long-period planets resulting from planet-planet scattering. This mechanism operates via dynamical friction with an outer disk (or with a cold planetesimal belt, i.e. \citealt{2018Eriksson}). The semi-major axes of such planets are limited by the size of the disk or belt, and simulations in \cite{2011Chatterjee} showed additional inward migration after circularization. Such planets would be on wide orbits, but of $\lesssim$100s of AU, mostly below the range we are considering here.

\cite{2006Fregeau} studied the cross-sections for interactions between a circular planet orbiting a host star and an intruder star of equal mass. They found a novel region of parameter space between the hard-soft boundary and the slow-fast boundary where there is about a $1/3$ probability that the planet will be stolen by the incoming star. This process occurs via an elastic (rather than head-on) collision of the two stars that --- from the perspective of the planet --- stops the intruder star and kicks out the original star. The other $2/3$ probability is ionization, in which the planet is stripped from its host star. This is the regime relevant to $100-1000$~AU planets ($v_{\rm orb} \simeq 1-3$~km/s) embedded in an open cluster ($v_\sigma \sim 1$~km/s). 

When a planet is swapped into an orbit around a different star (of equal mass), \cite{2006Fregeau} calculates its semi-major axis on average increases by $\sim$25~\%, and its new eccentricity is distributed thermally ($f(e)=2e$). What \cite{2006Fregeau} did not calculate is the cross-section for preserving the planet around its own host, yet changing the planet's eccentricity by a certain amount. Since we are presuming a more highly eccentric planet ($[1-e] \ll 1$, due to planet-planet scattering) initially around the first host star, then the orbit around the new host star will tend to be more circular, and thus whatever inner planetary system the new star has might be undisturbed by it (depending, of course, on the outer edge of that system). In our simulation, the fate of the planets was only determined with respect to the initial host star; it is possible but unknown if some of the planets that became unbound from their host star due to flybys in fact became bound to the flyby star. Finally, Fregeau's model considered equal-mass stars, which may be required for the elastic encounter to trade a planet \citep{2010Levison}. Such a coincidence in flyby stars is rare, given the broadness of the stellar initial mass function.

The similar model of \cite{2012Reipurth} tracks the ``dynamical unfolding'' of triple-star orbits, as the gaseous core that gives them birth evaporates. There, passing stars are considered as somewhat of a nuisance, whereas we treat them a vital mechanism for raising the periastron. We found this mechanism rather ineffective for 10 $M_J$ planets; it is likely even less effective for stars. Recently, \cite{2019DeRosa} proposed an interior binary HD 106906 kicked out a planet, but that planet was saved by a flying-by star (candidates were identified with \textit{Gaia}). Since we found that even massive planets would not efficiently create exo-Oort planets, an interior binary star would be even less likely as the initial scatterer in our mechanism. 

\cite{2006Gladman} say rogue planets may account for high perihelion scattered-disk objects such as Sedna. Maybe scattering among planets, with a quickly-lost middle planet, can do the same in exoplanetary systems. The idea was made much more specific in a suite of numerical experiments by \cite{2018Silsbee}. It deposits planets at a factor of 3-10 beyond the initial distribution of planets, at least for our Solar System's architecture.

The change of eccentricity due to an outer body has been studied in a number of contexts. \cite{1996Heggie} calculated the eccentricity change in a binary star that results from a flyby of a star in a cluster setting. The analytic expressions for eccentricity excitation, starting at zero eccentricity and in the limit that the flyby is slow, was derived by \cite{2001Kobayashi}. A similar calculation, but starting from an eccentricity of near unity, was done by \cite{2012KatzDong}. The latter's scientific question was whether the periastron of an inner binary in a triple can be changed significantly in one orbit of that binary. Thus it is very similar to our question, though they wanted to lower the periastron, and we want to raise it. Since they assumed the outer body was bound to the inner binary, it must be moving much slower than the inner binary components, so they were able to approximate the outer body as fixed at a given location. This approximation is the opposite extreme of the usual Oort-cloud approximation (the impulse approximation) in which the intruder star flies by much faster than the orbit of the comet. We are treating flybys in which the passage can be comparable to the orbit of the planet, so the details of the calculation may be different.

Simulations of planetary systems in stellar clusters have been done. \cite{2015Zheng} modeled single planet systems in evolving stellar clusters with varying semi-major axes of the planet. They found a sharp decrease in the survival rate of planets with $a$ $\gtrsim$ 2000 AU. The planets are on circular orbits for the entire lifetime of the cluster, which would produce a more destructive effect than predicted in our mechanism. \cite{2013Hao} modeled multi-planet systems in a stellar cluster. They found that planet-planet interactions are an important factor in understanding the impact of a dense stellar environment. Their planets showed eccentricity and semi-major axis growth over 100 Myr in a cluster, consistent with our expectations. Their cluster is of constant density over 100 Myr, which increases the number of close stellar interactions and reduces the survival rate of wide-orbit planets.

\subsection{Current and Potential Observations}

The obvious candidate for observing this population of wide-orbit planets directly requires imaging. The giant planets involved in planet-planet scattering will emit in the infrared and are expected to be observable around young stars. Recent observational surveys for wide-orbit planets have determined an upper frequency limit of 9\% for giant planets (0.5-13 $M_J$) at 100-1000 AU \citep{2016Durkan} and 3\% for giant planets (1-13 $M_J$) at 1000-5000 AU \citep{2018Baron}, rather larger than our predictions but not contradictory.

The current population of known wide-orbit planets is quite small, with only 10 confirmed planets with mass $<$13 $M_J$. The lowest mass of these is $\sim$6 $M_J$ (Ross 458 c; \citealt{2010Burgasser}). Planets of such high masses are very inefficiently formed via the Oort mechanism in our model, and we do not expect that any of the currently-known wide-orbit giant planets are formed via this pathway. Improved direct imaging detection limits, with sensitivities to lower masses and older star systems, could lead to detections of an Oort planet population.

Potentially, microlensing could infer this population and do so to much lower masses, with sensitivity to free-floating or loosely-bound planets down to Earth-mass objects \citep{2018Mroz}. \cite{2011Sumi} reported a large population of free-floating Jupiter-mass objects (1.8 objects per main sequence star) and, combined with direct imaging constraints on nearby potential hosts, determined that more than 75\% of these objects are either unbound or bound to stars further than 500 AU. A larger survey by \cite{2017Mroz} did not find evidence of this excess of Jupiter-mass objects but did find indications of Earth-mass and super-Earth-mass free-floating planets.

The systems with a loosely bound planet by our mechanism are the same systems that bear signatures at $<$5 AU of planet-planet scattering, the primary signature being giant planets with large eccentricity \citep{2008Juric,2008Chatterjee,2010Cumming}. If a system with far-out planets is found, a concerted search closer to the star is warranted, e.g., by closer-in imaging \citep{2016Bryan} or by radial velocity.

\subsection{Future Work}

This work focused on reproducing the Oort mechanism for giant planets and estimating a frequency that such planets might be saved rather than ejected for sun-like systems. More physically accurate simulations across a wider range of expected system properties, done for a more statistically robust number of simulations, could provide better predictions for the efficiency of the process as well as the expected properties of Oort planets. This more in-depth evaluation of Oort planets is deferred to future work.

Oort planets can act as external perturbers to the inner systems. In cases where the semi-major axis of the stranded planet lies beyond $\sim$3000 AU, over the lifetime of the star it may be torqued back into the inner system by Galactic tides and passing stars \citep{1986Heisler}, stimulating a new epoch of scattering. Hence Oort planets may become exo-Nemesis planets. The outcome of the instability generated by this long-timescale fuse awaits future work.

\acknowledgments

We acknowledge support of grant NASA-NNX17AB93G through NASA's Exoplanet Research Program. Simulations in this paper made use of the \texttt{REBOUND} code which can be downloaded freely at\\ http://github.com/hannorein/rebound.

\software{\texttt{REBOUND}, \texttt{Forecaster}
          }

\appendix
\section{Correcting for Variable Density}

\begin{figure}[ht]
\centering
\includegraphics[width=.5\linewidth]{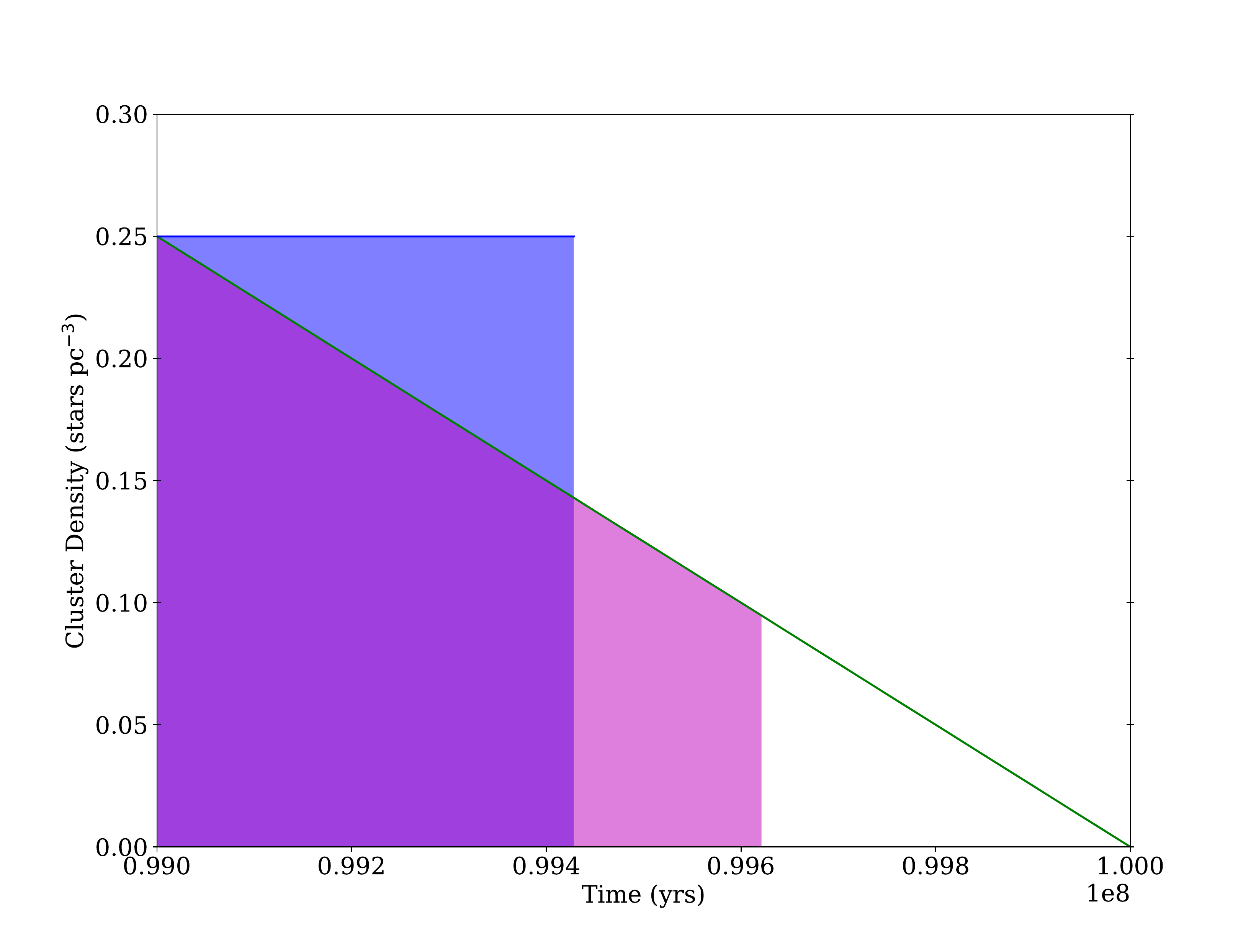}
\caption{Calculating the time delay between flyby stars. The green line shows the constantly decreasing density over the lifetime of the cluster. The width of the blue rectangle is the uncorrected wait time, assuming a constant density. The magenta trapezoid has the same area as the blue rectangle, and its width is the corrected wait time.}
\label{fig:dens_time}
\end{figure}

In determining the wait time between flyby stars, the fact that density is decreasing over time must be taken into account. An initial uncorrected wait time, $\tau_{o}$, is calculated using equation \ref{pass_rate}, assuming that the density is constant at some initial density, $n_{o}$. An example considering a flyby star during the last 1\% of a cluster's life is shown as a blue rectangle in Figure~\ref{fig:dens_time}, where $n_{init}=25\: \text{stars}\, \text{pc}^{-3}$. The additional delay due to the decaying density, assuming $\dot{n}$ is a negative constant over the life of the cluster, is found by calculating corrected wait time, $\tau$, that has the equivalent area underneath the density curve, as shown by the magenta trapezoid in Figure~\ref{fig:dens_time}.

\begin{equation}
n_{o}\tau_{o} = \tau(n_{o} + \frac{1}{2}\dot{n}\tau)
\end{equation}

Solving for $\tau$ leads to

\begin{equation}
\tau = \frac{-n_{o}\pm \sqrt{n_{o}^{2}+2n_{o}\dot{n}\tau_{o}}}{\dot{n}},
\end{equation}

where the positive root is the corrected wait time.

This process is iterated over the lifetime of the cluster, with $n_{o}$ for each new flyby taken to be the density at the time of the most recent flyby and a new random $\tau_{o}$ generated until $\tau$ implies a wait time beyond the age of the cluster.

\section{Probability Density Function of $b_{min}$}

Consider $x_i$ drawn from a uniform distribution between 0 and 1. The probability that $x_i$ is greater than $x$ is given by $1-x$. After $N$ such draws, reordering the results by index number, the probability that $x_o$ (the smallest number drawn) is greater than $x$ is given by

\begin{equation}
p(x_o>x)=\prod_{i=1}^{N}p(x_i>x) = (1-x)^N.
\end{equation}

Therefore the probability that $x_o$ is smaller than $x$ is given by

\begin{equation}
p(x_o<x) = 1-(1-x)^N.
\end{equation}

The probability density of $x$ being the smallest number after N draws is then computed as the derivative of $p(x_o<x)$.

\begin{equation}
p(x) = \left | \frac{dp(x_o<x)}{dx} \right | = N(1-x)^{N-1}
\end{equation}

This can be expanded to consider $f(x)=2x$ for $x$ drawn from a uniform distribution between 0 and 1. This is the same as drawing $x^2$ uniformly between 0 and 1, which is what we do for $x=\frac{b}{b_{max}}$. Now the probability that $x_i$ is greater than $x$ is given by integrating $f(x)$ from $x$ to 1. 

\begin{equation}
p(x_i>x)=\int_{x}^{1}f(x')dx' = 1-x^2
\end{equation}

The same logic as before can be applied to find the probability of $x_o$ being less than $x$.

\begin{equation}
p(x_o<x)=1-\prod_{i=1}^{N}p(x_i>x) = 1-(1-x^2)^N
\end{equation}

Finally, taking the absolute value of the derivative with respect to x, the probability density of $x$ being the smallest number after N draws is given by

\begin{equation}
p(x)=2xN(1-x^2)^{N-1}.
\end{equation}

Expanding further to a number $x^2$ drawn uniformly between 0 and $b_{max}^2$, we introduce a normalization factor $A$ and integrate over the interval to normalize:

\begin{equation}
\int_{0}^{b_{max}}A2xN(b_{max}^2-x^2)^{N-1}dx = 1.
\end{equation}

We find $A=b_{max}^{-2N}$ and therefore the probability density function can be written as

\begin{equation}\label{x_pdf}
p(x)=2b_{max}^{-2}xN(1-\frac{x^2}{b_{max}^2})^{N-1}.
\end{equation}

This is the same as equation \ref{bmin_pdf} where $x=b_{min}$.
Now, as we consider the limit where $N \rightarrow \infty$ and $b_{max}\rightarrow \infty$, we can establish a constant $d=\frac{b_{max}^2}{N}$. Equation \ref{x_pdf} can then be rewritten as

\begin{equation}\label{x_pdf2}
p(b_{min})=\frac{2b_{min}}{d}(1-\frac{b_{min}^2}{dN})^{N-1}
\end{equation}

Given the limit definition of Euler's formula,

\begin{equation}
\lim_{n\rightarrow \infty}(1+\frac{x}{n})^n = e^x,
\end{equation}

we can rewrite \ref{x_pdf2} as

\begin{equation}
p(b_{min})=\frac{2b_{min}}{d}e^{-\frac{b_{min}^2}{d}}.
\end{equation}

The probability density function of a Rayleigh distribution is 

\begin{equation}
f(x;\sigma )=\frac{x}{\sigma^2}e^{-x^2/(2\sigma^2)},
\end{equation}

so the probability density function of $b_{min}$ is a Rayleigh distribution with $\sigma=\sqrt{\frac{b_{max}^2}{2N}}$.

This derivation was inspired by the technique of finding the distribution of the highest peak among many in a periodogram \cite[Section IIIa]{1982Scargle}. A branch of statistics known as \emph{record statistics} (e.g., \citealt{1952Chandler}) can be put to service to find the distribution of the $N$th ``recent closest passage,'' though in practice we resorted to numerical draws to determine the distribution used for the efficiency calculation and illustrated in Figure~\ref{fig:all_bmin}.

\bibliography{ms} \bibliographystyle{aasjournal}

\begin{thebibliography}{}
\expandafter\ifx\csname natexlab\endcsname\relax\def\natexlab#1{#1}\fi
\providecommand{\url}[1]{\href{#1}{#1}}

\bibitem[{{Adams}(2010)}]{2010Adams}
{Adams}, F.~C. 2010, \araa, 48, 47

\bibitem[{{Andrews} {et~al.}(2009){Andrews}, {Wilner}, {Hughes}, {Qi}, \&
  {Dullemond}}]{2009Andrews}
{Andrews}, S.~M., {Wilner}, D.~J., {Hughes}, A.~M., {Qi}, C., \& {Dullemond},
  C.~P. 2009, \apj, 700, 1502

\bibitem[{{Baron} {et~al.}(2018){Baron}, {Artigau}, {Rameau}, {Lafreni{\`e}re},
  {Gagn{\'e}}, {Malo}, {Albert}, {Naud}, {Doyon}, {Janson}, {Delorme}, \&
  {Beichman}}]{2018Baron}
{Baron}, F., {Artigau}, {\'E}., {Rameau}, J., {et~al.} 2018, ArXiv e-prints,
  arXiv:1807.08799

\bibitem[{{Baruteau} {et~al.}(2011){Baruteau}, {Meru}, \&
  {Paardekooper}}]{2011Baruteau}
{Baruteau}, C., {Meru}, F., \& {Paardekooper}, S.-J. 2011, \mnras, 416, 1971

\bibitem[{{Batygin} \& {Brown}(2016)}]{2016Batygin}
{Batygin}, K., \& {Brown}, M.~E. 2016, \aj, 151, 22

\bibitem[{{Batygin} {et~al.}(2012){Batygin}, {Brown}, \& {Betts}}]{2012Batygin}
{Batygin}, K., {Brown}, M.~E., \& {Betts}, H. 2012, \apjl, 744, L3

\bibitem[{{Bowler}(2016)}]{2016Bowler}
{Bowler}, B.~P. 2016, \pasp, 128, 102001

\bibitem[{{Brasser} {et~al.}(2006){Brasser}, {Duncan}, \&
  {Levison}}]{2006Brasser}
{Brasser}, R., {Duncan}, M.~J., \& {Levison}, H.~F. 2006, \icarus, 184, 59

\bibitem[{{Brown} {et~al.}(2015){Brown}, {Bannister}, {Schmidt}, {Drake},
  {Djorgovski}, {Graham}, {Mahabal}, {Donalek}, {Larson}, {Christensen},
  {Beshore}, \& {McNaught}}]{2015Brown}
{Brown}, M.~E., {Bannister}, M.~T., {Schmidt}, B.~P., {et~al.} 2015, \aj, 149,
  69

\bibitem[{{Bryan} {et~al.}(2016){Bryan}, {Bowler}, {Knutson}, {Kraus},
  {Hinkley}, {Mawet}, {Nielsen}, \& {Blunt}}]{2016Bryan}
{Bryan}, M.~L., {Bowler}, B.~P., {Knutson}, H.~A., {et~al.} 2016, \apj, 827,
  100

\bibitem[{{Burgasser} {et~al.}(2010){Burgasser}, {Simcoe}, {Bochanski},
  {Saumon}, {Mamajek}, {Cushing}, {Marley}, {McMurtry}, {Pipher}, \&
  {Forrest}}]{2010Burgasser}
{Burgasser}, A.~J., {Simcoe}, R.~A., {Bochanski}, J.~J., {et~al.} 2010, \apj,
  725, 1405

\bibitem[{{Chambers} {et~al.}(1996){Chambers}, {Wetherill}, \&
  {Boss}}]{1996Chambers}
{Chambers}, J.~E., {Wetherill}, G.~W., \& {Boss}, A.~P. 1996, \icarus, 119, 261

\bibitem[{Chandler(1952)}]{1952Chandler}
Chandler, K.~N. 1952, Journal of the Royal Statistical Society. Series B
  (Methodological), 14, 220.
\newblock \url{http://www.jstor.org/stable/2983870}

\bibitem[{{Chatterjee} {et~al.}(2008){Chatterjee}, {Ford}, {Matsumura}, \&
  {Rasio}}]{2008Chatterjee}
{Chatterjee}, S., {Ford}, E.~B., {Matsumura}, S., \& {Rasio}, F.~A. 2008, \apj,
  686, 580

\bibitem[{{Chatterjee} {et~al.}(2011){Chatterjee}, {Ford}, \&
  {Rasio}}]{2011Chatterjee}
{Chatterjee}, S., {Ford}, E.~B., \& {Rasio}, F.~A. 2011, in IAU Symposium, Vol.
  276, The Astrophysics of Planetary Systems: Formation, Structure, and
  Dynamical Evolution, ed. A.~{Sozzetti}, M.~G. {Lattanzi}, \& A.~P. {Boss},
  225--229

\bibitem[{{Chen} \& {Kipping}(2017)}]{2017Chen}
{Chen}, J., \& {Kipping}, D. 2017, \apj, 834, 17

\bibitem[{{Cumming}(2010)}]{2010Cumming}
{Cumming}, A. 2010, {Statistical Distribution of Exoplanets}, ed. S.~{Seager},
  191--214

\bibitem[{{Dones} {et~al.}(2004){Dones}, {Weissman}, {Levison}, \&
  {Duncan}}]{2004Dones}
{Dones}, L., {Weissman}, P.~R., {Levison}, H.~F., \& {Duncan}, M.~J. 2004, in
  Astronomical Society of the Pacific Conference Series, Vol. 323, Star
  Formation in the Interstellar Medium: In Honor of David Hollenbach, ed.
  D.~{Johnstone}, F.~C. {Adams}, D.~N.~C. {Lin}, D.~A. {Neufeeld}, \& E.~C.
  {Ostriker}, 371

\bibitem[{{Durkan} {et~al.}(2016){Durkan}, {Janson}, \& {Carson}}]{2016Durkan}
{Durkan}, S., {Janson}, M., \& {Carson}, J.~C. 2016, \apj, 824, 58

\bibitem[{{Elmegreen}(2018)}]{2018Elmegreen}
{Elmegreen}, B.~G. 2018, \apj, 869, 119

\bibitem[{{Eriksson} {et~al.}(2018){Eriksson}, {Mustill}, \&
  {Johansen}}]{2018Eriksson}
{Eriksson}, L.~E.~J., {Mustill}, A.~J., \& {Johansen}, A. 2018, \mnras, 475,
  4609

\bibitem[{{Fern{\'a}ndez} \& {Brunini}(2000)}]{2000Fernandez}
{Fern{\'a}ndez}, J.~A., \& {Brunini}, A. 2000, \icarus, 145, 580

\bibitem[{{Fienga} {et~al.}(2016){Fienga}, {Laskar}, {Manche}, \&
  {Gastineau}}]{2016Fienga}
{Fienga}, A., {Laskar}, J., {Manche}, H., \& {Gastineau}, M. 2016, \aap, 587,
  L8

\bibitem[{{Ford} \& {Rasio}(2008)}]{2008Ford}
{Ford}, E.~B., \& {Rasio}, F.~A. 2008, \apj, 686, 621

\bibitem[{{Fregeau} {et~al.}(2006){Fregeau}, {Chatterjee}, \&
  {Rasio}}]{2006Fregeau}
{Fregeau}, J.~M., {Chatterjee}, S., \& {Rasio}, F.~A. 2006, \apj, 640, 1086

\bibitem[{{Gaudi} \& {Bloom}(2005)}]{2005Gaudi}
{Gaudi}, B.~S., \& {Bloom}, J.~S. 2005, \apj, 635, 711

\bibitem[{{Gladman} \& {Chan}(2006)}]{2006Gladman}
{Gladman}, B., \& {Chan}, C. 2006, \apjl, 643, L135

\bibitem[{{Hao} {et~al.}(2013){Hao}, {Kouwenhoven}, \& {Spurzem}}]{2013Hao}
{Hao}, W., {Kouwenhoven}, M.~B.~N., \& {Spurzem}, R. 2013, \mnras, 433, 867

\bibitem[{{Heggie} \& {Rasio}(1996)}]{1996Heggie}
{Heggie}, D.~C., \& {Rasio}, F.~A. 1996, \mnras, 282, 1064

\bibitem[{{Heisler} \& {Tremaine}(1986)}]{1986Heisler}
{Heisler}, J., \& {Tremaine}, S. 1986, \icarus, 65, 13

\bibitem[{{Henon}(1972)}]{1972Henon}
{Henon}, M. 1972, \aap, 19, 488

\bibitem[{{Juri{\'c}} \& {Tremaine}(2008)}]{2008Juric}
{Juri{\'c}}, M., \& {Tremaine}, S. 2008, \apj, 686, 603

\bibitem[{{Katz} \& {Dong}(2012)}]{2012KatzDong}
{Katz}, B., \& {Dong}, S. 2012, ArXiv e-prints, arXiv:1211.4584

\bibitem[{{Kobayashi} \& {Ida}(2001)}]{2001Kobayashi}
{Kobayashi}, H., \& {Ida}, S. 2001, \icarus, 153, 416

\bibitem[{{Lada} \& {Lada}(2003)}]{2003Lada}
{Lada}, C.~J., \& {Lada}, E.~A. 2003, \araa, 41, 57

\bibitem[{{Levison} {et~al.}(2010){Levison}, {Duncan}, {Brasser}, \&
  {Kaufmann}}]{2010Levison}
{Levison}, H.~F., {Duncan}, M.~J., {Brasser}, R., \& {Kaufmann}, D.~E. 2010,
  Science, 329, 187

\bibitem[{{Morbidelli} \& {Levison}(2004)}]{2004Morbidelli}
{Morbidelli}, A., \& {Levison}, H.~F. 2004, \aj, 128, 2564

\bibitem[{{Mr{\'o}z} {et~al.}(2017){Mr{\'o}z}, {Udalski}, {Skowron}, {Poleski},
  {Koz{\l}owski}, {Szyma{\'n}ski}, {Soszy{\'n}ski}, {Wyrzykowski},
  {Pietrukowicz}, {Ulaczyk}, {Skowron}, \& {Pawlak}}]{2017Mroz}
{Mr{\'o}z}, P., {Udalski}, A., {Skowron}, J., {et~al.} 2017, \nat, 548, 183

\bibitem[{{Mr{\'o}z} {et~al.}(2018){Mr{\'o}z}, {Ryu}, {Skowron}, {Udalski},
  {Gould}, {Szyma{\'n}ski}, {Soszy{\'n}ski}, {Poleski}, {Pietrukowicz},
  {Koz{\l}owski}, {Pawlak}, {Ulaczyk}, {OGLE Collaboration}, {Albrow}, {Chung},
  {Jung}, {Han}, {Hwang}, {Shin}, {Yee}, {Zhu}, {Cha}, {Kim}, {Kim}, {Kim},
  {Lee}, {Lee}, {Lee}, {Park}, {Pogge}, \& {KMTNet Collaboration}}]{2018Mroz}
{Mr{\'o}z}, P., {Ryu}, Y.-H., {Skowron}, J., {et~al.} 2018, \aj, 155, 121

\bibitem[{{Murray-Clay}(2010)}]{2010Murray-Clay}
{Murray-Clay}, R.~A. 2010, in Astronomical Society of the Pacific Conference
  Series, Vol. 432, New Horizons in Astronomy: Frank N. Bash Symposium 2009,
  ed. L.~M. {Stanford}, J.~D. {Green}, L.~{Hao}, \& Y.~{Mao}, 98

\bibitem[{{Nesvorn{\'y}}(2018)}]{2018Nesvorny}
{Nesvorn{\'y}}, D. 2018, \araa, 56, 137

\bibitem[{{Oort}(1950)}]{1950Oort}
{Oort}, J.~H. 1950, \bain, 11, 91

\bibitem[{{Perets} \& {Kouwenhoven}(2012)}]{2012Perets}
{Perets}, H.~B., \& {Kouwenhoven}, M.~B.~N. 2012, \apj, 750, 83

\bibitem[{{Proszkow} \& {Adams}(2009)}]{2009Proszkow}
{Proszkow}, E.-M., \& {Adams}, F.~C. 2009, \apjs, 185, 486

\bibitem[{{Rafikov}(2005)}]{2005Rafikov}
{Rafikov}, R.~R. 2005, \apjl, 621, L69

\bibitem[{{Rein} \& {Liu}(2012)}]{2012Rein}
{Rein}, H., \& {Liu}, S.-F. 2012, \aap, 537, A128

\bibitem[{{Rein} \& {Spiegel}(2015)}]{2015Rein}
{Rein}, H., \& {Spiegel}, D.~S. 2015, \mnras, 446, 1424

\bibitem[{{Reipurth} \& {Mikkola}(2012)}]{2012Reipurth}
{Reipurth}, B., \& {Mikkola}, S. 2012, \nat, 492, 221

\bibitem[{Rosa \& Kalas(2019)}]{2019DeRosa}
Rosa, R. J.~D., \& Kalas, P. 2019, The Astronomical Journal, 157, 125

\bibitem[{{Scargle}(1982)}]{1982Scargle}
{Scargle}, J.~D. 1982, \apj, 263, 835

\bibitem[{{Schwamb} {et~al.}(2010){Schwamb}, {Brown}, {Rabinowitz}, \&
  {Ragozzine}}]{2010Schwamb}
{Schwamb}, M.~E., {Brown}, M.~E., {Rabinowitz}, D.~L., \& {Ragozzine}, D. 2010,
  \apj, 720, 1691

\bibitem[{{Seidelmann} \& {Harrington}(1988)}]{1988Seidelmann}
{Seidelmann}, P.~K., \& {Harrington}, R.~S. 1988, Celestial Mechanics, 43, 55

\bibitem[{{Silsbee} \& {Tremaine}(2018)}]{2018Silsbee}
{Silsbee}, K., \& {Tremaine}, S. 2018, \aj, 155, 75

\bibitem[{{Standish}(1993)}]{1993Standish}
{Standish}, E.~M. 1993, \aj, 105, 2000

\bibitem[{{Stone} {et~al.}(2015){Stone}, {Metzger}, \& {Loeb}}]{2015Stone}
{Stone}, N., {Metzger}, B.~D., \& {Loeb}, A. 2015, \mnras, 448, 188

\bibitem[{{Sumi} {et~al.}(2011){Sumi}, {Kamiya}, {Bennett}, {Bond}, {Abe},
  {Botzler}, {Fukui}, {Furusawa}, {Hearnshaw}, {Itow}, {Kilmartin}, {Korpela},
  {Lin}, {Ling}, {Masuda}, {Matsubara}, {Miyake}, {Motomura}, {Muraki},
  {Nagaya}, {Nakamura}, {Ohnishi}, {Okumura}, {Perrott}, {Rattenbury}, {Saito},
  {Sako}, {Sullivan}, {Sweatman}, {Tristram}, {Udalski}, {Szyma{\'n}ski},
  {Kubiak}, {Pietrzy{\'n}ski}, {Poleski}, {Soszy{\'n}ski}, {Wyrzykowski},
  {Ulaczyk}, \& {Microlensing Observations in Astrophysics (MOA)
  Collaboration}}]{2011Sumi}
{Sumi}, T., {Kamiya}, K., {Bennett}, D.~P., {et~al.} 2011, \nat, 473, 349

\bibitem[{{Tremaine}(1993)}]{1993Tremaine}
{Tremaine}, S. 1993, in Astronomical Society of the Pacific Conference Series,
  Vol.~36, Planets Around Pulsars, ed. J.~A. {Phillips}, S.~E. {Thorsett}, \&
  S.~R. {Kulkarni}, 335--344

\bibitem[{{Trujillo} \& {Sheppard}(2014)}]{2014Trujillo}
{Trujillo}, C.~A., \& {Sheppard}, S.~S. 2014, \nat, 507, 471

\bibitem[{{van Elteren} {et~al.}(2019){van Elteren}, {Portegies Zwart},
  {Pelupessy}, {Cai}, \& {McMillan}}]{2019vanElteren}
{van Elteren}, A., {Portegies Zwart}, S., {Pelupessy}, I., {Cai}, M., \&
  {McMillan}, S. 2019, arXiv e-prints, arXiv:1902.04652

\bibitem[{{Veras} \& {Armitage}(2004)}]{2004Veras}
{Veras}, D., \& {Armitage}, P.~J. 2004, \mnras, 347, 613

\bibitem[{{Veras} {et~al.}(2009){Veras}, {Crepp}, \& {Ford}}]{2009Veras}
{Veras}, D., {Crepp}, J.~R., \& {Ford}, E.~B. 2009, \apj, 696, 1600

\bibitem[{{Veras} {et~al.}(2014){Veras}, {Shannon}, \&
  {G{\"a}nsicke}}]{2014Veras}
{Veras}, D., {Shannon}, A., \& {G{\"a}nsicke}, B.~T. 2014, \mnras, 445, 4175

\bibitem[{{Volk} \& {Malhotra}(2017)}]{2017Volk}
{Volk}, K., \& {Malhotra}, R. 2017, \aj, 154, 62

\bibitem[{{Ward} \& {Kruijssen}(2018)}]{2018Ward}
{Ward}, J.~L., \& {Kruijssen}, J.~M.~D. 2018, \mnras, 475, 5659

\bibitem[{{Whitmire} \& {Jackson}(1984)}]{1984Whitmire}
{Whitmire}, D.~P., \& {Jackson}, A.~A. 1984, \nat, 308, 713

\bibitem[{{Wright} \& {Mamajek}(2018)}]{2018Wright}
{Wright}, N.~J., \& {Mamajek}, E.~E. 2018, \mnras, 476, 381

\bibitem[{{Zakamska} \& {Tremaine}(2005)}]{2005Zakamska}
{Zakamska}, N.~L., \& {Tremaine}, S. 2005, \aj, 130, 1939

\bibitem[{{Zheng} {et~al.}(2015){Zheng}, {Kouwenhoven}, \& {Wang}}]{2015Zheng}
{Zheng}, X., {Kouwenhoven}, M.~B.~N., \& {Wang}, L. 2015, \mnras, 453, 2759

\end{thebibliography}

\end{document}